\documentclass[12pt]{article}

\renewcommand{\title}[1]{
\begin{center} \Large \bf #1 \end{center}
}

\renewcommand{\author}[3]{
 \begin{center} #1 \\ \ \\
  {\it #2} \\ \ \\
  {\texttt{#3}}
 \end{center}
}

\usepackage{amssymb}
\usepackage{amsmath}

\usepackage[dvips]{graphicx}

\font\mybb=msbm10 at 12pt

\def\bb#1{\hbox{\mybb#1}}

\begin{document}

\begin{titlepage}

\begin{flushleft}
September 2004
\end{flushleft}

\begin{flushright}
KSTS/RR-04/006 \\
OCHA-PP-234 \ \ \ \ \ 
\end{flushright}

\title{Recovery of Full $N=1$ Supersymmetry in Non(anti-)commutative
 Superspace} 

\author
{Akifumi Sako${}^{\dagger}$ and Toshiya Suzuki${}^{\ast} {}^{\S}$ \\ \ }
{${}^{\dagger}$ Department of Mathematics, Faculty of Science and
 Technology, Keio University\\
3-14-1 Hiyoshi, Kohoku-ku, Yokohama 223-8522, Japan\\ \ \\
${}^{\ast}$ Department of Physics, Faculty of Science, Ochanomizu University\\
and\\
${}^{\S}$ Institute of Humanities and Sciences, Ochanomizu University\\
2-1-1 Otsuka, Bunkyo-ku, Tokyo 112-8610, Japan\\ \ }
{${}^{\dagger}$ sako@math.keio.ac.jp \\
${}^{\ast}$ tsuzuki@phys.ocha.ac.jp }

\vspace{1cm}

\abstract{
\noindent
We investigate SUSY of Wess-Zumino models in non(anti-)commutative
 Euclidean superspaces. Non(anti-)commutative deformations break 1/2
 SUSY, then non(anti-)commutative Wess-Zumino models do not have
 full SUSY in general. However, we can recover full SUSY at specific coupling
 constants satisfying some relations. We give a general way to construct
 full SUSY non(anti-)commutative Wess-Zumino models. For a some example,
 we investigate quantum corrections and $\beta$-functions behavior.
}

%

\end{titlepage}

\section{Introduction}
Non(anti-)commutative superspaces have attracted much
interest \cite{FerraraO} - \cite{GrTa}.
One of significant features of non(anti-)commutative 
field theories is 1/2 SUSY.
Non(anti-)commutative theories do not have full SUSY, but one half of SUSY
is preserved in many cases.
Non(anti-)commutative theories are constructed by several deformations
of usual (anti-)commutative SUSY theories.
We will treat 2 types of the deformations 
given by SUSY $*$ product and non-SUSY
$\star$ product. (These definitions will be given in the next section.)
The SUSY $*$ product does not break the SUSY algebra, but SUSY $*$
products of chiral superfields are not chiral usually, then 1/2 SUSY is
broken (see, for example, \cite{Ferrara}). 
On the other hand, non-SUSY $\star$ products of chiral superfields are
chiral superfields, but the SUSY algebra itself is broken 
as we saw in \cite{Seiberg} and so on.
\\

In this article, we will show that 
tuning of coupling constants regains full SUSY
for not only SUSY $*$ deformation cases but also
non-SUSY $\star$ deformation cases where the normal SUSY algebra is broken.
In non(anti-)commutative theories, all lagrangians, algebras and
transformation laws should be defined by using $*$ (or $\star$ etc.) products.
Since the $*$ (or $\star$ etc.) product often breaks SUSY, 
it is non-trivial question to ask that
one can define the SUSY algebra in non(anti-)commutative superspaces 
or
there exist non(anti-)commutative full SUSY lagrangians.
At first we will propose a general method to
construct the full SUSY Wess-Zumino models in non(anti-)commutative
superspaces. 
Using this procedure, we will understand that full SUSY recovers at
specific values of the coupling constants.
After that we will observe the quantum effect around the
supersymmetry recovering points.
The quantum corrections will be calculated 
and $\beta$-functions will be determined
at one-loop order.
In some phases, full SUSY is stable in the IR limit.

\section{Conventions}
We study the Wess-Zumino models on 4-dim 
Euclidean superspace and use the convention of
\cite{WeBa}.
Non(anti-)deformation is 
given by several ways. 
For example, in \cite{Ferrara}, we can see the
systematic explanation of the deformations.
So we use the notation of \cite{Ferrara}.\\

The left covariant derivative is identical to
the ordinary supersymmetric covariant derivative,
\begin{equation}
\overrightarrow{D} \Phi := D \Phi \ .
\label{lD}
\end{equation}
On the other hand, the right covariant derivative is defined through the
following relation.
\begin{equation}
\Phi \overleftarrow{D} := (-1)^{p_{D} (p_{\Phi} + 1)} \overrightarrow{D} \Phi \ ,\end{equation}
where $p_A$ is parity of $A$ i.e. $p_A=0$ for bosonic $A$
and $p_B=1$ for fermionic $B$.
With this setting, for some superfields $\Phi$ and $\Psi$ the SUSY $*$
product is defined by 
\begin{eqnarray}
\Phi * \Psi &:=& \Phi \cdot \Psi + P^{\alpha \beta} \Phi \overleftarrow{D}_{\alpha} \overrightarrow{D}_{\beta} \Psi + \frac{1}{4} det P \Phi \overleftarrow{D}^{2} \overrightarrow{D}^{2} \Psi \nonumber \\
 &=& \Phi \cdot \Psi - P^{\alpha \beta} D_{\alpha} \Phi D_{\beta} \Psi - \frac{1}{4} det P D^{2} \Phi D^{2} \Psi \ .
\label{SP4}
\end{eqnarray}
This $*$ product is naively extended to the product of operators.
For arbitrary operators (or superfields) $O_1$ and $O_2$,
\begin{eqnarray}
O_1 * O_2 
 &=& O_1 \cdot O_2 +(-)^{{p_D}{(p_{O1}+1)}}
  P^{\alpha \beta}\left[ D_{\alpha} , O_1 \} [D_{\beta} ,O_2 \right\}
  \nonumber \\
&& -\frac{1}{4} det P 
 \epsilon^{\beta \alpha} 
 \left[ D_{\alpha} , \left[ D_{\beta} , O_1 \right\} \right\}
 \epsilon^{\delta \gamma} 
 \left[D_{\gamma} , \left[D_{\delta} , O_2 \right\} \right\}
  \ ,
\label{SpO}
\end{eqnarray}
where $\left[ A , B \right\} := AB - (-)^{p_A p_B} BA$.

We introduce another typical non(anti-)commutative deformation
with using not $D$ but $Q$. 
The left SUSY generating operator action is identical to
the ordinary action
\begin{equation}
\overrightarrow{Q} \Phi := Q \Phi \ .
\label{lQ}
\end{equation}
On the other hand, the right action is defined by
\begin{equation}
\Phi \overleftarrow{Q} := (-1)^{p_{Q} (p_{\Phi} + 1)} \overrightarrow{Q} \Phi \ .\end{equation}
With this setting, the non-SUSY $\star$ product is defined by
\begin{eqnarray}
\Phi \star \Psi &:=& \Phi \cdot \Psi + P^{\alpha \beta} \Phi \overleftarrow{Q}_{\alpha} \overrightarrow{Q}_{\beta} \Psi + \frac{1}{4} det P \Phi \overleftarrow{Q}^{2} \overrightarrow{Q}^{2} \Psi \nonumber \\
 &=& \Phi \cdot \Psi - P^{\alpha \beta} Q_{\alpha} \Phi Q_{\beta} \Psi - \frac{1}{4} det P Q^{2} \Phi Q^{2} \Psi \ .
\label{nonSUSYstar}
\end{eqnarray}

\section{Full SUSY in Non(anti-)commutative Field Theory}
In this section, we propose a general procedure to build full
SUSY lagrangians in non(anti-)commutative ${\bb R}^4$ from 
1/2 SUSY lagrangians.\\

We will begin by considering the SUSY algebra determined by 
the SUSY $*$ product.
\begin{eqnarray}
\{ Q_{\alpha} , Q_{\beta}\}_* &=& 
\{ \bar{Q}_{\dot{\alpha}}, \bar{Q}_{\dot{\beta}}\}_* =0\\
\{ Q_\alpha,  \bar{Q}_{\dot{\beta}} \}_* 
&=& 2 \sigma^{\mu}_{\alpha \dot{\beta}} P_{\mu} = - 2i 
\sigma^\mu_{\alpha \dot{\beta}} \partial_\mu \\
\left[ P_\mu ,  Q_{\alpha} \right]_* &=& 
\left[ P_{\mu}, \bar{Q}_{\dot{\alpha}} \right]_*=0\\
\left[M_{\mu \nu}, Q_\alpha \right]_* &=& 
-(i\sigma_{\mu \nu} )_{\alpha}^{\beta}Q_{\beta} \\
\left[ M_{\mu \nu}, \bar{Q}^{\dot{\alpha}} \right]_* &=&
 -
(i\bar{\sigma}_{\mu \nu} )_{\dot{\alpha}}^{\dot{\beta}}\bar{Q}^{\dot{\beta}} ,
\end{eqnarray}
where $\{ A ,B\}_*:= A*B+B*A$ and $[A,B]_*:= A*B - B*A$. The 
definition of SUSY $*$ product (\ref{SP4}) uses the covariant derivative $D_{\alpha}$, 
and all generators of supersymmetric Poincare group commute with
$D_{\alpha}$, therefore the above SUSY algebra is trivial.
We call the symmetry generated by the above algebra
super-$*$ symmetry, i.e. a super-$*$ symmetric lagrangian $L$ satisfies
that
\begin{eqnarray}
\int d^4 x  \ Q_{\alpha} * L =0 \ , \ 
 \int d^4 x \  \bar{Q}_{\dot{\alpha}} * L = 0 \ .
\end{eqnarray}
Note that the super-$*$ symmetry is defined in the non(anti-)commutative superspace.
\\

Let us construct full SUSY Wess-Zumino like models in the non(anti-)commutative
Euclidean space that is deformed by the SUSY $*$ product.
Let $\Phi$ be a chiral super field.
Let $\int d^4 \theta V^0(\Phi ; *) \bar{\theta}^2$ 
be a 1/2 SUSY invariant lagrangian,
where the $V^0(\Phi ; *) $ is a some polynomial in $\Phi$ and its derivative
$\partial_{\mu} \Phi$ , $D_{\alpha} \Phi$ and so on.
A general 1/2 SUSY lagrangian is given from the elements of the 
1/2 SUSY ring discussed in \cite{SaSu}.
In $V^0(\Phi ; *)$, all products 
are $*$ products, but $V^0(\Phi ; *)$ is possible to be expressed by 
usual products by using (\ref{SP4}).
Let $\tilde{V}^0(\Phi ; \cdot )$
denote the usual product representation of $V^0(\Phi ; *) $ i.e.
$V^0(\Phi ; *) = \tilde{V}^0(\Phi ; \cdot )$.
We can regard $\tilde{V}^0(\Phi ; \cdot )$ as the sum of supersymmetric 
lagrangian $\tilde{V}^0_s(\Phi ; \cdot )$ and 1/2 SUSY lagrangian
$-\tilde{V}^0_{1/2}(\Phi ; \cdot )$:
\begin{equation}
\tilde{V}^0(\Phi ; \cdot )= \tilde{V}^0_s(\Phi ; \cdot )
-\tilde{V}^0_{1/2}(\Phi ; \cdot ) \ ,
\end{equation}
where $\tilde{V}^0_s(\Phi ; \cdot )$ and $\tilde{V}^0_{1/2}(\Phi ; \cdot
)$ satisfy 
\begin{eqnarray}
\int d^4x d^4\theta  \bar{\theta}^2 
Q_{\alpha} \tilde{V}^0_s(\Phi ; \cdot )= 0 \ & , & \
\int d^4x d^4\theta  \bar{\theta}^2 
\bar{Q}_{\dot{\alpha}} \tilde{V}^0_s(\Phi ; \cdot )= 0 \\
\int d^4x d^4\theta  \bar{\theta}^2 
Q_{\alpha} \tilde{V}^0_{1/2}(\Phi ; \cdot )= 0 \ & , & \
\int d^4x d^4\theta  \bar{\theta}^2 
\bar{Q}_{\dot{\alpha}} \tilde{V}^0_{1/2}(\Phi ; \cdot )\neq 0.
\end{eqnarray}
Next step, we introduce a new lagrangian $\int d^4 \theta {\bar \theta}^2
V^1(\Phi ; *)$ by
\begin{eqnarray}
\int d^4 \theta \{ V^1(\Phi ; *) \}\bar{\theta}^2
:= \int d^4 \theta \{ V^0(\Phi ; *) 
+ {V}^0_{1/2}(\Phi ; * )\} \bar{\theta}^2 \ ,  \label{1-lagrangian}
\end{eqnarray}
where ${V}^0_{1/2}(\Phi ; * )$ is 
$\tilde{V}^0_{1/2}(\Phi ; \cdot )$ deformed by replacing all usual
products with SUSY $*$ products.
{}From (\ref{SP4}),
\begin{eqnarray}
{V}^0_{1/2}(\Phi ; * )&=& \tilde{V}^0_{1/2}(\Phi ; \cdot )
+ (\makebox{higher order of $P^{\alpha \beta}$} ) \nonumber \\
&:=& \tilde{V}^0_{1/2}(\Phi ; \cdot ) +
 \tilde{V}^1_{def}(\Phi ,P^{\alpha \beta} ; \cdot ). \label{v1}
\end{eqnarray}
In general, we can divide 
$\tilde{V}^1_{def}(\Phi ,P^{\alpha \beta} ; \cdot )$ 
into a full SUSY part $\tilde{V}^1_{s}(\Phi ,P^{\alpha \beta} ; \cdot )$
and a 1/2 SUSY part $\tilde{V}^1_{1/2}(\Phi ,P^{\alpha \beta} ; \cdot )$:
\begin{equation}
\tilde{V}^1_{def}(\Phi ,P^{\alpha \beta}; \cdot )= \tilde{V}^1_s(\Phi ,P^{\alpha \beta}; \cdot ) -\tilde{V}^1_{1/2}(\Phi ,P^{\alpha \beta}; \cdot ),
\label{v1def}
\end{equation}
where $\tilde{V}^1_s(\Phi ,P^{\alpha \beta}; \cdot )$ and
$\tilde{V}^1_{1/2}(\Phi ,P^{\alpha \beta}; \cdot )$ satisfy 
\begin{eqnarray}
\int d^4x d^4\theta  \bar{\theta}^2 
Q_{\alpha} \tilde{V}^1_s(\Phi ,P^{\alpha \beta}; \cdot )= 0 \ & , & \ 
\int d^4x d^4\theta  \bar{\theta}^2 
\bar{Q}_{\dot{\alpha}} \tilde{V}^1_s(\Phi ,P^{\alpha \beta}; \cdot )= 0 \\
\int d^4x d^4\theta  \bar{\theta}^2 
Q_{\alpha} \tilde{V}^1_{1/2}(\Phi ,P^{\alpha \beta}; \cdot )= 0 \ & , & \
\int d^4x d^4\theta  \bar{\theta}^2 
\bar{Q}_{\dot{\alpha}} \tilde{V}^1_{1/2}(\Phi ,P^{\alpha \beta}; \cdot )\neq 0.
\end{eqnarray}
{}From (\ref{1-lagrangian}),(\ref{v1}) and (\ref{v1def}), the lagrangian is 
\begin{eqnarray}
\int d^4 \theta \{ V^1(\Phi ; *) \}\bar{\theta}^2
= \int d^4 \theta \bar{\theta}^2 \{ [\tilde{V}^0_s(\Phi ; \cdot )+
\tilde{V}^1_s(\Phi ,P^{\alpha \beta}; \cdot )]
-\tilde{V}^1_{1/2}(\Phi ,P^{\alpha \beta}; \cdot ) \} . \label{1-lagrangian2}
\end{eqnarray}
Note that $\tilde{V}^0_s(\Phi ; \cdot )+
\tilde{V}^1_s(\Phi ,P^{\alpha \beta}; \cdot )$ in the right hand side
is full supersymmetric.
We can duplicate the process from (\ref{1-lagrangian}) to 
(\ref{1-lagrangian2}) to eliminate the 1/2 SUSY terms.
The $n$-th process is as follows.
The $n$-th lagrangian is 
\begin{eqnarray}
\int d^4 \theta \{ V^n(\Phi ; *) \}\bar{\theta}^2
:= \int d^4 \theta \{ V^{n-1}(\Phi ; *) 
+ {V}^{n-1}_{1/2}(\Phi ; * )\} \bar{\theta}^2 \ ,  \label{n-lagrangian}
\end{eqnarray}
where ${V}^{n-1}_{1/2}(\Phi ; * )$ is 
$\tilde{V}^{n-1}_{1/2}(\Phi ; \cdot )$ deformed by replacing 
usual products by SUSY $*$ products.
{}From (\ref{SP4}),
\begin{eqnarray}
{V}^{n-1}_{1/2}(\Phi ; * )&=& 
\tilde{V}^{n-1}_{1/2}(\Phi ,P^{\alpha \beta}; \cdot )
+ (\makebox{higher order of $P^{\alpha \beta}$} ) \nonumber \\
&:=& \tilde{V}^{n-1}_{1/2}(\Phi ; \cdot ) +
 \tilde{V}^n_{def}(\Phi ,P^{\alpha \beta} ; \cdot ). \label{vn}
\end{eqnarray}
We divide 
$\tilde{V}^n_{def}(\Phi ,P^{\alpha \beta} ; \cdot )$ 
into a full SUSY part $\tilde{V}^n_{s}(\Phi ,P^{\alpha \beta} ; \cdot )$
and a 1/2 SUSY part $\tilde{V}^n_{1/2}(\Phi ,P^{\alpha \beta} ; \cdot )$.
Using this, the $n$-th lagrangian is given as
\begin{eqnarray}
\int d^4 \theta \{ V^n(\Phi ; *) \}\bar{\theta}^2
= \int d^4 \theta \bar{\theta}^2 &\big\{&[ \tilde{V}^0_s(\Phi ; \cdot )+
\tilde{V}^1_s(\Phi ,P^{\alpha \beta}; \cdot )
+\cdots +
\tilde{V}^n_s(\Phi ,P^{\alpha \beta}; \cdot ) ] \nonumber \\
&-&\tilde{V}^n_{1/2}(\Phi ,P^{\alpha \beta}; \cdot ) \big\} . 
\label{n-lagrangian2}
\end{eqnarray}

The key point is these processes get over at finite rotation.
Because the deformation of (\ref{n-lagrangian}) makes 
$\tilde{V}^n_{1/2}(\Phi ,P^{\alpha \beta}; \cdot )$
be higher order terms of $P^{\alpha \beta}$.
In proportion to the square root of the power of 
$P^{\alpha \beta}$, the number of $D_{\alpha}$ in 
$\tilde{V}^n_{1/2}(\Phi ,P^{\alpha \beta}; \cdot )$
increases. Since $D_{\alpha}D_{\beta}D_{\gamma}=0$,
there is a finite number $N$ such that 
$\tilde{V}^{N}_{1/2}(\Phi ,P^{\alpha \beta}; \cdot )=0$.

{}Then we get the full SUSY (super-$*$ symmetric) lagrangian
\begin{eqnarray}
\int d^4 \theta \{ V^N(\Phi ; *) \}\bar{\theta}^2 &=& \int d^4 \theta {\bar \theta}^2 \{ V^0(\Phi ; *) + [ V^0_{1/2}(\Phi ; *) + \cdots + V^{N-1}_{1/2}(\Phi ; *) ] \} \label{N-lagrangian2} \\
 &=& \int d^4 \theta \bar{\theta}^2 \{ \tilde{V}^0_s(\Phi ; \cdot )+
\tilde{V}^1_s(\Phi ,P^{\alpha \beta}; \cdot )
+ \cdots + \tilde{V}^N_{s}(\Phi ,P^{\alpha \beta}; \cdot ) \} . \nonumber
\end{eqnarray}

We will take examples to illustrate the above method to
construct full SUSY lagrangians.\\

The first example is 
\begin{eqnarray}
S = \int d^4 x d^2 \theta d^2 {\bar \theta} 
{\bar \Phi}* \Phi + \int d^4 x d^2 {\bar \theta} \frac{\bar m}{2} 
{\bar \Phi}* {\bar \Phi}+ \int d^4 x d^2 \theta \{ \frac{m}{2}\Phi*\Phi
 + g_{0*} \Phi*\Phi*\Phi \}.
\end{eqnarray}
The quadratic terms are not deformed by the SUSY $*$ product.
Then the only $\Phi*\Phi*\Phi $ should be modified to get the full SUSY action.
Let us follow the above instruction in the condition
$\Phi*\Phi*\Phi =V^0(\Phi ; *) $.
Rewriting this as  
\begin{eqnarray}
\Phi*\Phi*\Phi=\Phi^3 - \frac{1}{4}\det P 
\Phi D^2 \Phi D^2 \Phi =\tilde{V}^0(\Phi ; \cdot ) \ ,
\end{eqnarray}
then $\tilde{V}_s^0(\Phi , \cdot ) =\Phi^3$ and 
$\tilde{V}_{1/2}^0(\Phi , \cdot )= \frac{1}{4}\det P 
\Phi D^2 \Phi D^2 \Phi$.
Thus,
\begin{eqnarray}
V_{1/2}^0(\Phi , * )= \frac{1}{4}\det P 
\Phi * D_*^2 * \Phi * D_*^2 * \Phi , \label{p3v2}
\end{eqnarray}
where $D_*^2:=D^{\alpha}*D_{\alpha}$. 
In the following, we often use that $D_*^2=D^2$, $D*\Phi=D\Phi$ etc. 
Using (\ref{p3v2}), a new lagrangian is introduced by
\begin{eqnarray}
\int d^4 \theta \bar{\theta}^2
\{ V^1( \Phi ; * ) \}&=& \int d^4 \theta \bar{\theta}^2
\{ \Phi*\Phi*\Phi + \frac{1}{4}\det P 
\Phi * D^2  \Phi * D^2 \Phi \} \nonumber \\
&=& \int d^4 \theta \bar{\theta}^2 \Phi^3 .
\end{eqnarray}
This is the super-$*$ symmetric (full SUSY) lagrangian.
This lagrangian is regarded as a special case of 1/2 SUSY
lagrangian
%
%
$\int d^4 \theta \bar{\theta}^2
\{g_{0*} \Phi*\Phi*\Phi + g_{1*}
\Phi * D^2  \Phi * D^2 \Phi \}$,
where $g_{0*}$ and $g_{1*}$ are
coupling constants. 
{}From this point of view, we can observe that tuning coupling constants, 
%
%
$$ g_{1*} / g_{0*} \rightarrow \frac{1}{4}\det P \ , $$
realizes full SUSY.
\\

The second example is the case of $\Phi_*^4 =V^0(\Phi ; *) $,
where $\Phi_*^n = \underbrace{\Phi * \cdots * \Phi}_n$. 
{}From 
\begin{eqnarray}
\Phi_*^4 = \Phi^4 -\frac{1}{4}\det P \Phi^2 D^2 \Phi D^2 \Phi 
-\frac{1}{4}\det P  D^2 \Phi D^2 \Phi^3
+\frac{1}{16} (\det P )^2 (D^2 \Phi)^4 \ , 
\end{eqnarray}
$\tilde{V}^0_s$ and $\tilde{V}^0_{1/2}$ are given by
\begin{eqnarray}
\tilde{V}^0_s(\Phi , \cdot )&=&\Phi^4  \nonumber \\
\tilde{V}^0_{1/2}(\Phi , \cdot )&=&
\frac{1}{4}\det P \Phi^2 D^2 \Phi D^2 \Phi 
+\frac{1}{4}\det P  D^2 \Phi D^2 \Phi^3
-\frac{1}{16} (\det P )^2 (D^2 \Phi)^4 \ . \nonumber 
\end{eqnarray}
Therefore, the modified lagrangian is given by
\begin{eqnarray}
&&\int d^4 \theta \bar{\theta}^2 {V}^1(\Phi , * )
=\int d^4 \theta \bar{\theta}^2
\{ {V}^0(\Phi , * ) + {V}^0_{1/2}(\Phi , * )\} \nonumber \\
&=& 
\int d^4 \theta \bar{\theta}^2 \{ \Phi^4_* 
+ [ \frac{1}{4}\det P \Phi^2_* D^2 \Phi D^2 \Phi 
+\frac{1}{4}\det P  D^2 \Phi D^2 \Phi^3_*
-\frac{1}{16} (\det P )^2 (D^2 \Phi)^4 ] \} \nonumber \\
&=&\int d^4 \theta \bar{\theta}^2 \ \{ \Phi^4 
- \frac{1}{8}(\det P)^2(D^2 \Phi)^4 \} \ \nonumber \\
&=&\int d^4 \theta \bar{\theta}^2 \ \ \Phi^4 .
\end{eqnarray}
This is what we want.
Again, we can regard this lagrangian is given 
by tuning of coupling constants of the 1/2 SUSY lagrangian.
\\

This method is extended easily to the case of non-SUSY $\star$ deformation.
However, the SUSY algebra for non-SUSY $\star$ product has to be
modified, because the $\star$ definition (\ref{nonSUSYstar})
uses $Q_{\alpha}$ and it does not commute with $\bar{Q}_{\dot{\alpha}}$.
So we replace $\bar{Q}_{\dot{\alpha}}$ by $\bar{\mathcal Q}_{\dot{\alpha}}$
defined by \footnote{This $\bar{\mathcal Q}_{\dot{\alpha}}$ first
appeared in \cite{Seiberg}.}
\begin{eqnarray}
\bar{\mathcal Q}_{\dot{\alpha}}
:= \bar{Q}_{\dot{\alpha}}+  P^{\alpha \beta}
 \{ Q_{\alpha} , \bar{Q}_{\dot{\alpha}} \} Q_{\beta} 
= \bar{Q}_{\dot{\alpha}}- 2 P^{\alpha \beta} ( i \sigma^{\mu}_{\alpha \dot{\alpha}} \partial_{\mu} ) Q_{\beta}.
\end{eqnarray}
{}From this definition, $\bar{\mathcal Q}_{\dot{\alpha}}$ satisfies
\begin{eqnarray}
\bar{\mathcal Q}_{\dot{\alpha}} \star \phi = 
\bar{Q}_{\dot{\alpha}} \phi ,
\end{eqnarray}
for arbitrary field $\phi=\phi(x,\theta,{\bar \theta})$.
Using $\bar{\mathcal Q}_{\dot{\alpha}}$, the super-$\star$ symmetry 
algebra is defined by
\begin{eqnarray}
\{ Q_{\alpha} , Q_{\beta}\}_\star &=& 
\{ \bar{{\mathcal Q}}_{\dot{\alpha}}, \bar{{\mathcal Q}}_{\dot{\beta}}\}_\star =0\\
\{ Q_\alpha,  \bar{{\mathcal Q}}_{\dot{\beta}} \}_\star 
&=& 2 \sigma^{\mu}_{\alpha \dot{\beta}} P_{\mu} = - 2i 
\sigma^\mu_{\alpha \dot{\beta}} \partial_\mu \\
\left[ P_\mu ,  Q_{\alpha} \right]_\star &=& 
\left[ P_{\mu}, \bar{{\mathcal Q}}_{\dot{\alpha}} \right]_\star=0\\
\left[M_{\mu \nu}, Q_\alpha \right]_\star &=& 
-(i\sigma_{\mu \nu} )_{\alpha}^{\beta}Q_{\beta} \\
\left[ M_{\mu \nu}, \bar{{\mathcal Q}}^{\dot{\alpha}} \right]_\star &=&
 -
(i\bar{\sigma}_{\mu \nu} )_{\dot{\alpha}}^{\dot{\beta}}\bar{{\mathcal Q}}^{\dot{\beta}} .
\end{eqnarray}
Here $\{ A ,B\}_{\star}:= A\star B+B\star A$ and 
$[A,B]_\star := A\star B - B\star A$.
Note that $\bar{\mathcal Q}_{\dot{\alpha}}$ is not derivative 
because it includes second derivative and 
does not satisfy the Leibniz rule.
So, we can not make a group from the above super-$\star$ symmetry algebra.
However, it is possible to make full super-$\star$ symmetric field theories, 
where the super-$\star$ symmetry is defined by invariance under
\begin{eqnarray}
\Phi &\rightarrow & \Phi + \zeta^{\alpha}Q_{\alpha} \star \Phi 
+\bar{\zeta}_{\dot{\alpha}}\bar{\mathcal Q}^{\dot{\alpha}} \star \Phi
= \Phi + \zeta^{\alpha}Q_{\alpha} \Phi +
\bar{\zeta}_{\dot{\alpha}}\bar{Q}^{\dot{\alpha}} \Phi
\ , \nonumber \\
\bar{\Phi} &\rightarrow & \bar{\Phi}+ \zeta^{\alpha}Q_{\alpha} \star \bar{\Phi}
+\bar{\zeta}_{\dot{\alpha}}\bar{\mathcal Q}^{\dot{\alpha}}
\star \bar{\Phi}
=\bar{\Phi}+ \zeta^{\alpha}Q_{\alpha} \bar{\Phi}
+\bar{\zeta}_{\dot{\alpha}} \bar{ Q}^{\dot{\alpha}}
\bar{\Phi}
\ , \nonumber 
\end{eqnarray}
where $\zeta$ and $\bar{\zeta}$ are fermionic spinor parameters.
This super-$\star$ symmetry is defined as 
a symmetry of non(anti-)commutative 
field theory.
As mentioned above, 
this transformation does not mean the 
super Poincare group transformation of the superspace,
because $\bar{\mathcal Q}_{\dot{\alpha}}$ is not
derivative and it does not make a group action,
in the sense of non(anti-)commutative theory.
Therefore, this symmetry is not a superspace symmetry but 
a symmetry defined by an infinitesimal field transformation.
However, arbitrary super-$\star$ symmetric lagrangians satisfy
\begin{eqnarray}
\int d^4 x \   Q_{\alpha} \star L = 0 \ , \ 
\int d^4 x  \ 
\bar{\mathcal Q}_{\dot{\alpha}} \star L = 0 . 
\end{eqnarray}
We can construct super-$\star$ symmetric 
lagrangians by the same way of above super-$*$ symmetric 
lagrangian construction.
\\

There are some comments.
The method in this section is formally extended to 
other 1/2 SUSY theories in non(anti-)commutative superspaces.
We expect, for example, 1/2 SUSY gauge theories are deformed to
full SUSY theories by this method.
However, there may be some problems on eliminating SUSY breaking terms.
There are no assurance that this process stop at finite rotation, for
some kind of theories like non-abelian gauge theories.
Also, it is not clear that the (deformed) gauge invariance is maintained in
this procedure.
Therefore, more detailed analyses are needed.

The super-$*$($\star$) symmetric action in non(anti-)commutative superspace
is equivalent to the normal SUSY action given by $V^0(\Phi;*(\star))$ replacing
$*$($\star$) with the usual multiplication, in above two examples.
It may be that there are equivalent non(anti-)commutative theories
to arbitrary SUSY theories in usual space.

In this article, we treat the $*$ and $\star$ deformed theories. There
are many kinds of non(anti-)commutative deformations, and it is known
that some deformations do not break any SUSY (see, for example,
\cite{Park}). In any case, when 
deformed products are expressed by a polynomial with finite terms, the
above method is valid to construct full SUSY non(anti-)commutative
lagrangians.
      
\section{1-loop Calculations}
In the previous section, we gave the prescription to obtain full SUSY
actions in non(anti-)commutative superspaces. 
In this section, we will investigate quantum effects \cite{BFR1}-\cite{BBP}, concentrating on
the $\Phi_*^3$ model for simplicity. \\

{}From the results of the previous section, we know that
\begin{eqnarray}
S_\cdot &=& \int d^4 x d^2 \theta d^2 {\bar \theta} {\bar \Phi} \Phi + \int d^4 x d^2 \theta d^2 {\bar \theta} \theta^2 \frac{\bar m}{2} {\bar \Phi}^2 \nonumber \\
 &+& \int d^4 x d^2 \theta d^2 {\bar \theta} {\bar \theta}^2 \{ \frac{m}{2}\Phi^2 + g_{0} \Phi^3 + g_{1} \Phi D^2 \Phi D^2 \Phi \},
\label{action-dot} 
\end{eqnarray}
and 
\begin{eqnarray}
S_* &=& \int d^4 x d^2 \theta d^2 {\bar \theta} {\bar \Phi} * \Phi + \int d^4 x d^2 \theta d^2 {\bar \theta} \theta^2 \frac{\bar m}{2} {\bar \Phi}_*^2 \nonumber \\
 &+& \int d^4 x d^2 \theta d^2 {\bar \theta} {\bar \theta}^2 \{ \frac{m}{2}\Phi_*^2 + g_{0*} \Phi_*^3 + g_{1*} \Phi * D_*^2 * \Phi * D_*^2 * \Phi \},
\label{action-star} 
\end{eqnarray}
are equivalent, if
\begin{eqnarray}
g_0 = g_{0*} \ , \ g_1 = g_{1*} - \frac{1}{4} det P \ g_{0*}.
\label{g-g*}
\end{eqnarray}
We will calculate 1-loop graphs and $\beta$-functions based on
(\ref{action-dot}),\footnote{
In \cite{GPR1}, $\beta$-functions based on slightly different action
were calculated at 2-loop level. Also, in \cite{AGS},
$\beta$-functions for the non(anti-)commutative gauge theory were
calculated at 1-loop level.
} and interpret the results in terms of (\ref{action-star}).

Before calculating 1-loop corrections, it is convenient to integrate out 
${\bar \Phi}$ in (\ref{action-dot}) \cite{TeYe,DGLVZ}.
\begin{equation}
S = \int d^4 x d^2 \theta d^2 {\bar \theta} {\bar \theta}^2 \{ \frac{1}{2} \Phi(m - \frac{\square}{\bar m})\Phi + g_{0} \Phi^3 + g_{1} \Phi D^2 \Phi D^2 \Phi \}.
\label{action1}
\end{equation}
Also we have to add a term proportional to $\Phi D^2 \Phi$ to
renormalize a 1-loop divergent graph (See Fig.1) \cite{BFR1,TeYe}. So we
take the following action as the starting point:
\begin{equation}
S = \int d^4 x d^2 \theta d^2 {\bar \theta} {\bar \theta}^2 \{ \frac{1}{2} \Phi_b(\frac{1}{4 \lambda_b}D^2 + m_b - \frac{\square}{\bar m_b})\Phi_b + g_{0b} \Phi_b^3 + g_{1b} \Phi_b D^2 \Phi_b D^2 \Phi_b \},
\label{action2}
\end{equation}
where the subscription $b$ denotes $\Phi_b$ etc. are bare quantities. In
\cite{BrFe,Romagnoni,LuRe,BeRe}, the renormalizability of (\ref{action2}) was
proved.

To renormalize 1-loop corrections, we define the following renormalized quantities:
\begin{eqnarray}
 & \Phi_b = \sqrt{Z_\Phi} (\Phi_r + \delta \Phi) \ \ , \ \ m_b = m_r + \delta m \ \ , \ \ {\bar m_b} = Z_{\bar m} Z_{\Phi} {\bar m_r} \ \ , & \nonumber \\
 & \lambda_b = Z_{\lambda} Z_{\Phi} \lambda_r \ \ , \ \ g_{0b} = Z_0 \sqrt{Z_\Phi}^{-3} g_{0r} \ \ , \ \ g_{1b} = Z_1 \sqrt{Z_\Phi}^{-3} g_{1r} \ \ , &
\label{z}
\end{eqnarray}
where
\begin{eqnarray}
 & Z_{\Phi} = 1 + Z_{\Phi}^{(1)} + ... \ \ , \ \ \delta \Phi = 0 + \delta \Phi^{(1)} + ... \ \ , & \nonumber \\ 
 & \delta m = 0 + \delta m^{(1)} + ... \ \ , \ \ Z_{\bar m} = 1 + Z_{\bar m}^{(1)} + ... \ \ , & \nonumber \\
 & Z_{\lambda} = 1 + Z_{\lambda}^{(1)} + ... \ \ , \ \ Z_{0} = 1 + Z_{0}^{(1)} + ... \ \ , \ \ Z_{1} = 1 + Z_{1}^{(1)} + ... & .
\label{z1}
\end{eqnarray}

1-loop calculations tell us
\begin{eqnarray}
 & \delta m = 0 \ \ , \ \ Z_{\bar m}^{(1)} = 0 \ \ , & \nonumber \\
 & Z_{0}^{(1)} = 0 \ \ , \ \ Z_{1}^{(1)} = 0 \ \ , &
\label{z1_0} \\
\end{eqnarray}
and
\begin{equation}
\frac{1}{2} Z_{\Phi}^{(1)} m_r + 3 g_{0r} \delta \Phi^{(1)} = 0 ,
\label{z1phi-z1delta}
\end{equation}
so we find $Z_\lambda$ and $\delta \Phi$ are independent.
They are determined by the renormalization for 1-loop divergent graphs,
Fig.1 and Fig.2.

\begin{center}
\begin{minipage}{7cm}
\begin{center}
\includegraphics[width=5cm,clip]{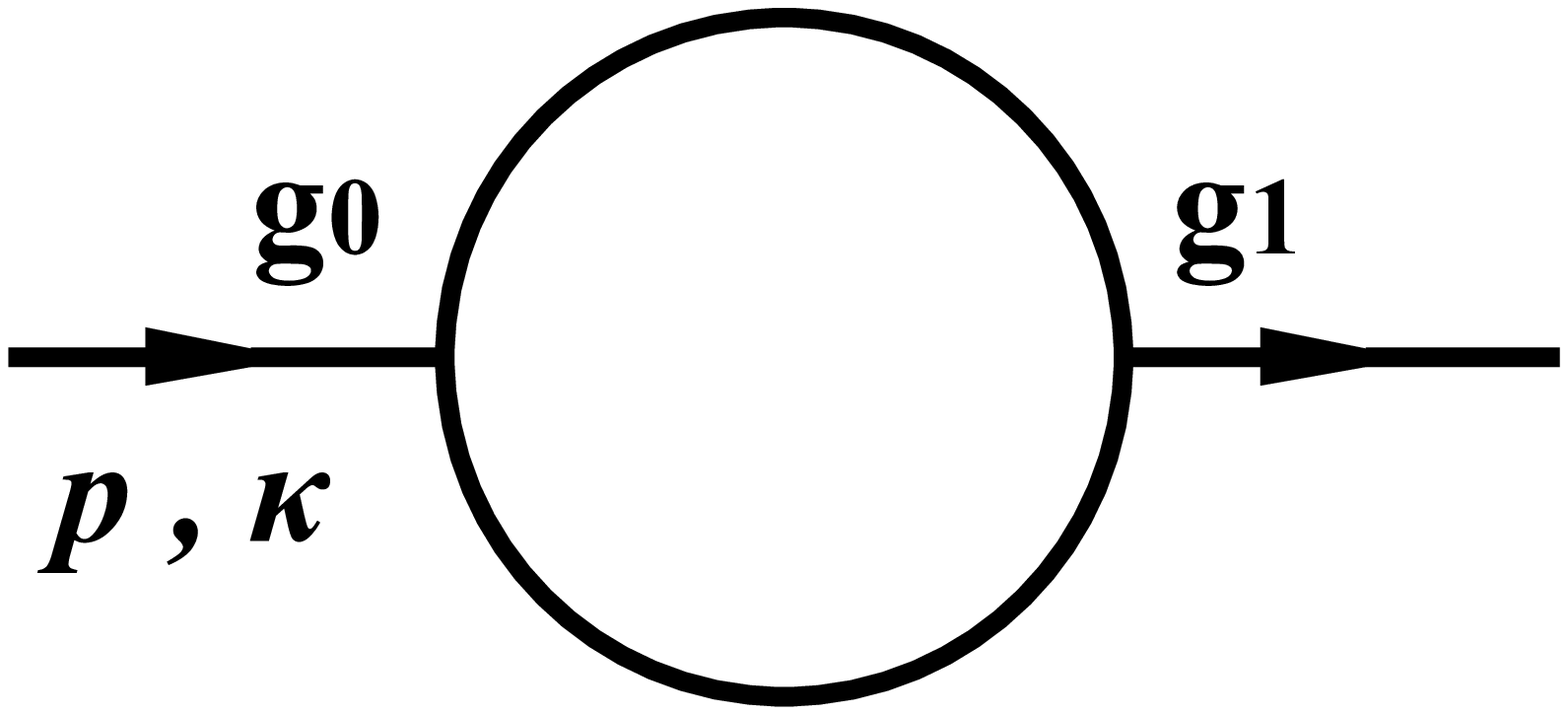} \\ \vspace{-3cm} Fig.1: 1-loop divergent
 graph contributing to $\frac{1}{4\lambda}\Phi D^2 \Phi$. 
\end{center}
\end{minipage}
\begin{minipage}{7cm}
\vspace{-1cm}
\begin{center}
\includegraphics[width=5cm,clip]{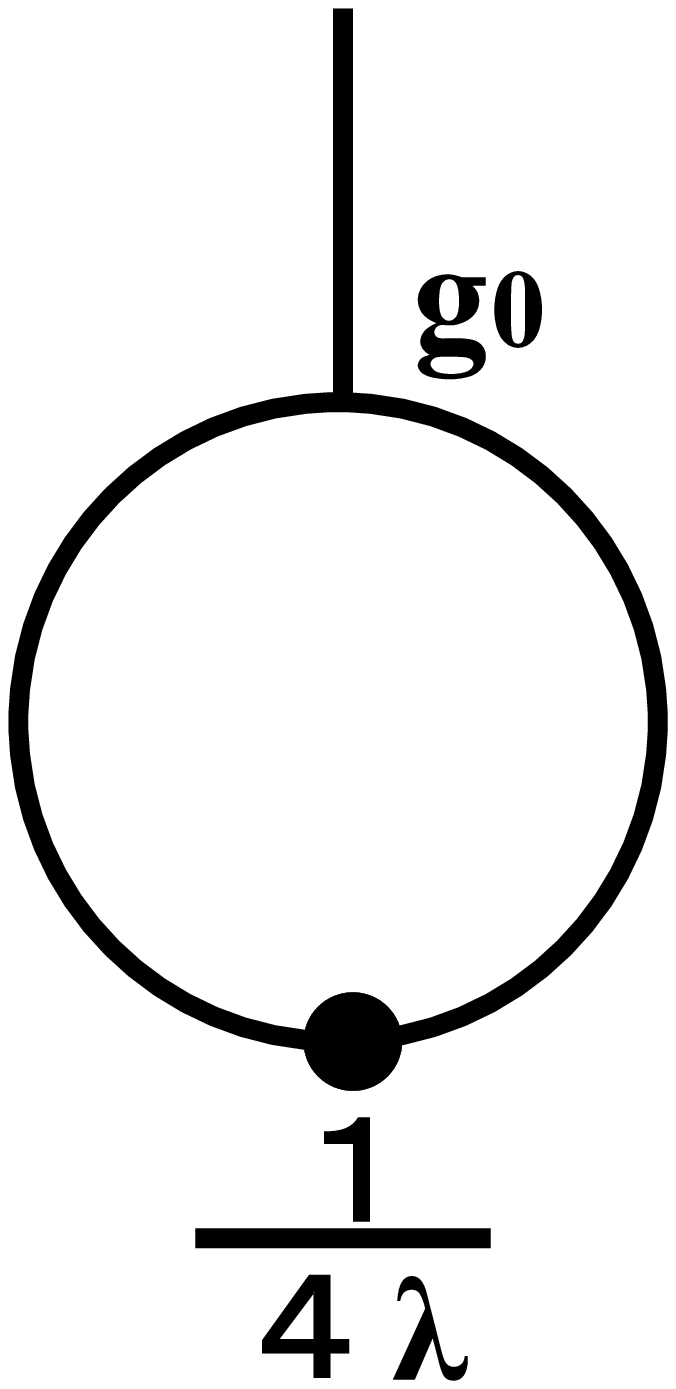} \\ \vspace{-2cm} Fig.2: 1-loop divergent
 tadpole graph canceled by $\delta \Phi$. 
\end{center}
\end{minipage}
\end{center}

It is worthwhile to comment on the role of $\frac{1}{4 \lambda}\Phi D^2
\Phi$ in (\ref{action2}). Firstly, as we mentioned above, this is
necessary to renormalize the 1-loop divergent graph, Fig.1. Secondly,
this term changes the propagator:
\begin{eqnarray}
\frac{\bar m}{m{\bar m} + p^2} &\Rightarrow& \frac{\bar m}{m{\bar m} + p^2 + \frac{\bar m}{4\lambda}\kappa^2} \ \ , \nonumber \\
 & & = \frac{\bar m}{m{\bar m} + p^2} -  \frac{\bar m}{m{\bar m} + p^2} \frac{\kappa^2}{4 \lambda} \frac{\bar m}{m{\bar m} + p^2} \ ,
\label{prop}
\end{eqnarray}
where $\kappa$ is the conjugate momentum of the fermionic coordinate
$\theta$.
This change yields the divergent tadpole graph, Fig.2, at the 1-loop
level. This non-vanishing tadpole causes the field shift $\delta \Phi$,
and the wave function renormalization through the relation
(\ref{z1phi-z1delta}). It is shown that the above field shift does not
lift the vacuum energy, then it does not break the (1/2) SUSY
\cite{BFR1,TeYe}.

We adopt the dimensional regularization, that is, we replace the
dimension $4$ by $n\in{\bb C}$, and the minimal subtraction
renormalization. So we introduce a scale parameter 
$\mu$ whose mass dimension is $1$ and redefine the renormalized
quantities by
\begin{eqnarray}
m_b &=& \mu {\tilde m_r},
\label{mbr} \\
{\bar m}_b &=& Z_{\Phi} \mu {\tilde {\bar m}}_r,
\label{bmbr} \\
\lambda_b &=& Z_{\lambda} Z_{\Phi} \lambda_r,
\label{lbr} \\
g_{0b} &=& Z_{\Phi}^{-\frac{3}{2}} \mu^{\frac{4-n}{2}} g_{0r},
\label{g0br} \\
g_{1b} &=& Z_{\Phi}^{-\frac{3}{2}} \mu^{\frac{-n}{2}} {\tilde g_{1r}}.
\label{g1br}
\end{eqnarray}
Here, ${\tilde m_r}$, ${\tilde {\bar m}}_r$, $\lambda_r$, $g_{0r}$ and
${\tilde g_{1r}}$ are dimensionless. 

Let us determine $Z_\lambda$. The contribution of Fig.1 is
\begin{equation}
\Gamma_{D^2}^{(2)} = - \frac{9 g_{0r} {\tilde g_{1r}} {\tilde {\bar m}}_r^2}{\pi^2} \kappa^2 \left( \frac{1}{\epsilon} + ... \right),
\label{div1}
\end{equation}
where $\epsilon = \frac{4-n}{2}$.
The divergent part of (\ref{div1}) is canceled by the counter term
$\frac{Z_{\lambda}^{(1)}}{4 \lambda_r} \Phi_r D^2 \Phi_r$,
so
\begin{equation}
Z_{\lambda}^{(1)} = \frac{1}{\epsilon} \left( -\frac{36 g_{0r} {\tilde g_{1r}} {\bar {\tilde m}}_r^2 \lambda_r}{\pi^2} \right).
\label{zl1}
\end{equation}

Now we turn to $\delta \Phi$ and $Z_\Phi$. The contribution of Fig.2 is
\begin{equation}
\Gamma^{(1)} = \frac{3 g_{0r} \mu^2 {\tilde {\bar m}}_r^2}{16 \pi^2 \lambda_r} \left( \frac{1}{\epsilon} + ... \right) .
\label{div2}
\end{equation}
The divergent part of (\ref{div2}) is canceled by the counter term $\mu {\tilde
m_r} \delta \Phi^{(1)}$. From (\ref{z1phi-z1delta}), we obtain 
\begin{equation}
Z_{\phi}^{(1)} = \frac{1}{\epsilon} \left( \frac{9 g_{0r}^2 {\tilde {\bar m}}_r^2}{8 \pi^2 {\tilde m}_r^2 \lambda_r} \right).
\label{zphi1}
\end{equation}

Using (\ref{zl1}) and (\ref{zphi1}), we calculate the $\beta$-functions 
defined by
\begin{eqnarray}  
 & \beta_0 = \frac{\partial g_{0r}}{\partial log \mu} \ \ , \ \ \beta_{\tilde 1} = \frac{\partial {\tilde g}_{1r}}{\partial log \mu} \ \ , \ \ \beta_{\lambda^{-1}} = \frac{\partial \lambda_r^{-1}}{\partial log \mu} \ \ , & \nonumber \\
 & \beta_{\tilde m} = \frac{\partial {\tilde m}_r}{\partial log \mu} \ \ , \ \ \beta_{\tilde {\bar m}} = \frac{\partial {\tilde {\bar m}}_r}{\partial log \mu}\ \ .
\label{defbeta}
\end{eqnarray}
The results are
\begin{eqnarray}
 & \beta_0 = - \frac{27 g_{0r}^3 \lambda_r^{-1} {\tilde {\bar m}}_r^2}{8 \pi^2 {\tilde m}_r^2} \ \ , \ \ \beta_{\tilde 1} = 2{\tilde g}_{1r} - \frac{27 g_{0r}^2 {\tilde g}_{1r} \lambda_r^{-1} {\tilde {\bar m}}_r^2}{8 \pi^2 {\tilde m}^2} \ \ , & \nonumber \\
 & \beta_{\lambda^{-1}} = - \frac{9 g_{0r}^2 \lambda_r^{-2} {\tilde {\bar m}}_r^2}{4 \pi^2 {\tilde m}_r^2} + \frac{72 g_{0r} {\tilde g}_{1r} {\tilde {\bar m}}_r^2}{\pi^2} \ \ , & \nonumber \\
 & \beta_{\tilde m} = - {\tilde m}_r \ \ , \ \ \beta_{\tilde {\bar m}} = - {\tilde {\bar m}}_r - \frac{9 g_{0r}^2 \lambda_r^{-1} {\tilde {\bar m}}_r^{3}}{4 \pi^2 {\tilde m}_r^2} \ \ . &
\label{beta}
\end{eqnarray}

\section{Discussions}
In this section, we argue the RG behavior of $\Phi_*^3$ model,
particularly, around points realizing full SUSY.
Notice that full SUSY is recovered when $g_1=0$ and
$\lambda^{-1}=0$. Naively, since $g_1$ has mass dimension $-2$, we
expect full SUSY is repaired in the IR limit. However, as we will see in
the following, the RG behavior of $\lambda^{-1}$ modifies this
speculation. \\

It makes things clear to classify situations according to the sign of
$\lambda^{-1}$. From (\ref{beta}),

\noindent
(i)$\lambda^{-1} > 0$ (Fig.3a-3c) \\
For $g_0$, $g_0 \rightarrow 0$ in the UV limit. For $\lambda^{-1}$, \\
$\bullet$ \ $\lambda^{-1} \rightarrow 0$ in IR, if \ $\{ g_0 > 0 \ , \ g_1
> \frac{\lambda^{-2}}{32 m^2} g_0 \}$ \ or \ $\{ g_0 < 0 \ , \ g_1 < \frac{\lambda^{-2}}{32 m^2} g_0 \}$, \\
$\bullet$ \ $\lambda^{-1} \rightarrow 0$ in UV, if \ $\{ g_0 > 0 \ , \ g_1
< \frac{\lambda^{-2}}{32 m^2} g_0 \}$ \ or \ $\{ g_0 < 0 \ , \ g_1 > \frac{\lambda^{-2}}{32 m^2} g_0 \}$.

\begin{center}
\begin{minipage}{5cm}
\vspace{-1cm}
\begin{center}
\includegraphics[width=5cm,clip]{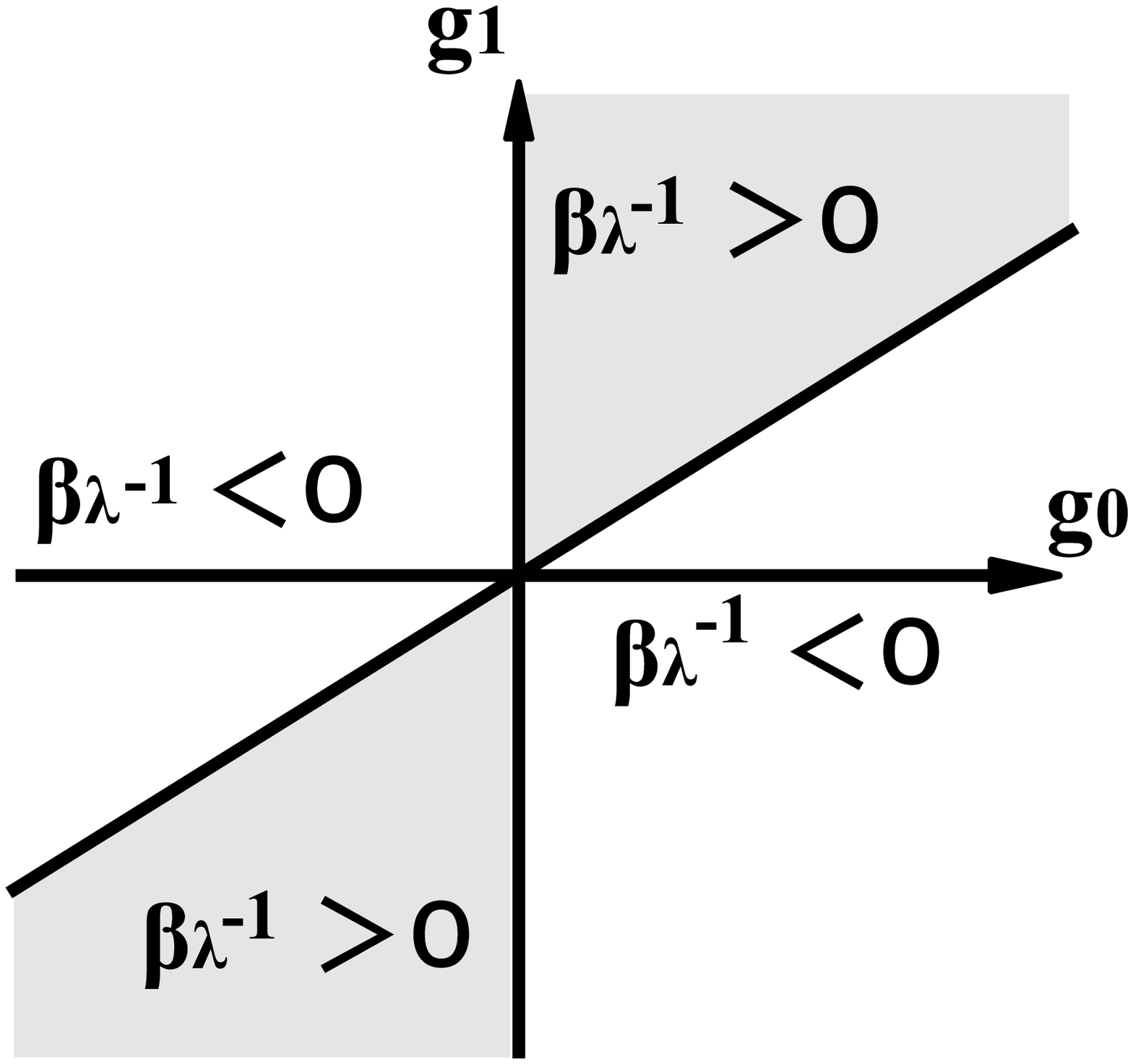} \\ \vspace{-2cm} Fig.3a:
 $\beta_{\lambda^{-1}}>0$ in the grayed region. The oblique line represents $g_1=\frac{\lambda^{-2}}{32m^2}g_1$.
\end{center}
\end{minipage} \ \
\begin{minipage}{5cm}
\vspace{-1cm}
\begin{center}
\includegraphics[width=5cm,clip]{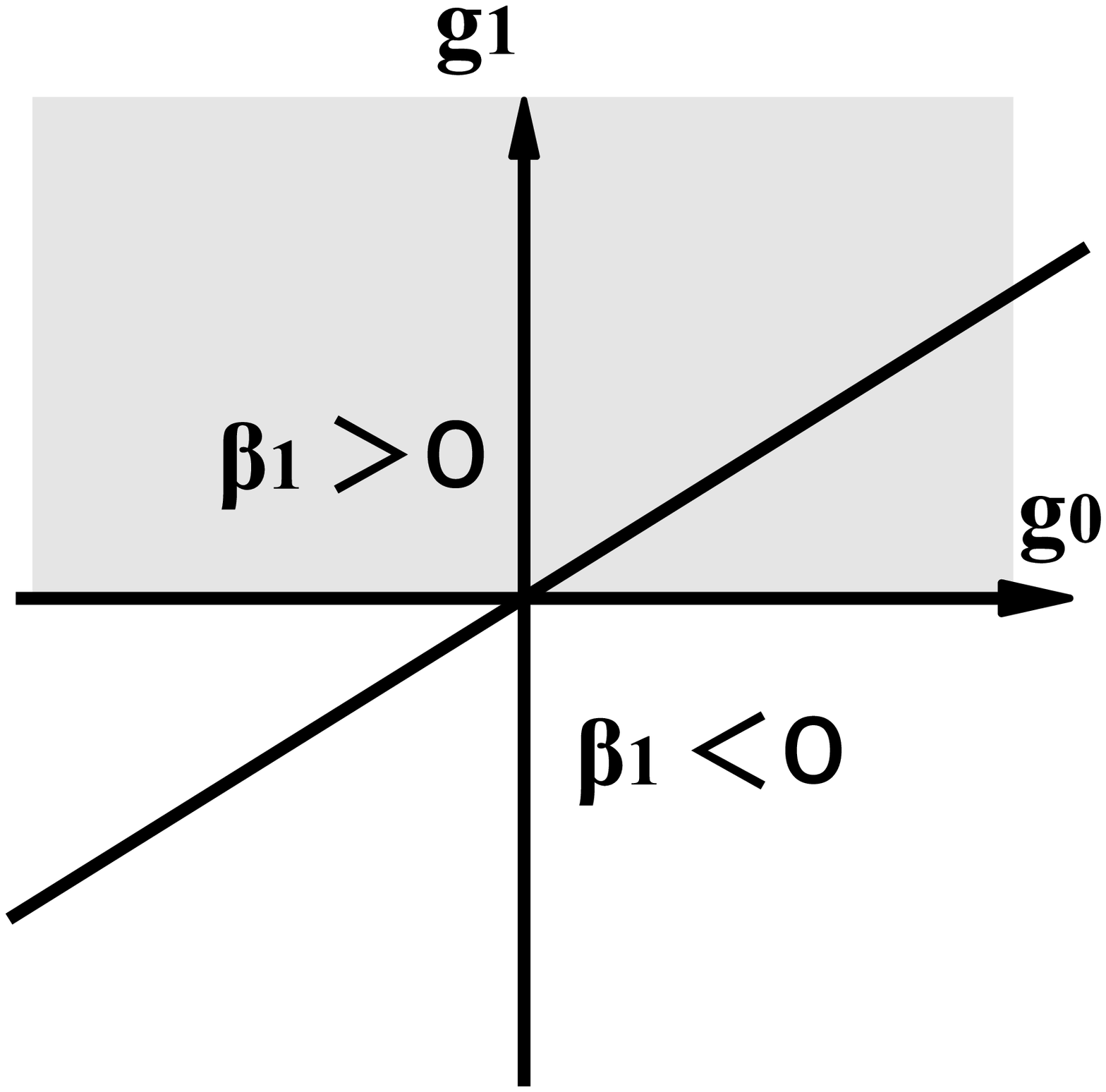} \\ \vspace{-2cm} Fig.3b:
 $\beta_{\tilde 1}>0$ in the grayed region. The oblique line
 represents $g_1=\frac{\lambda^{-2}}{32m^2}g_1$.
\end{center}
\end{minipage} \ \
\begin{minipage}{5cm}
\vspace{-1cm}
\begin{center}
\includegraphics[width=5cm,clip]{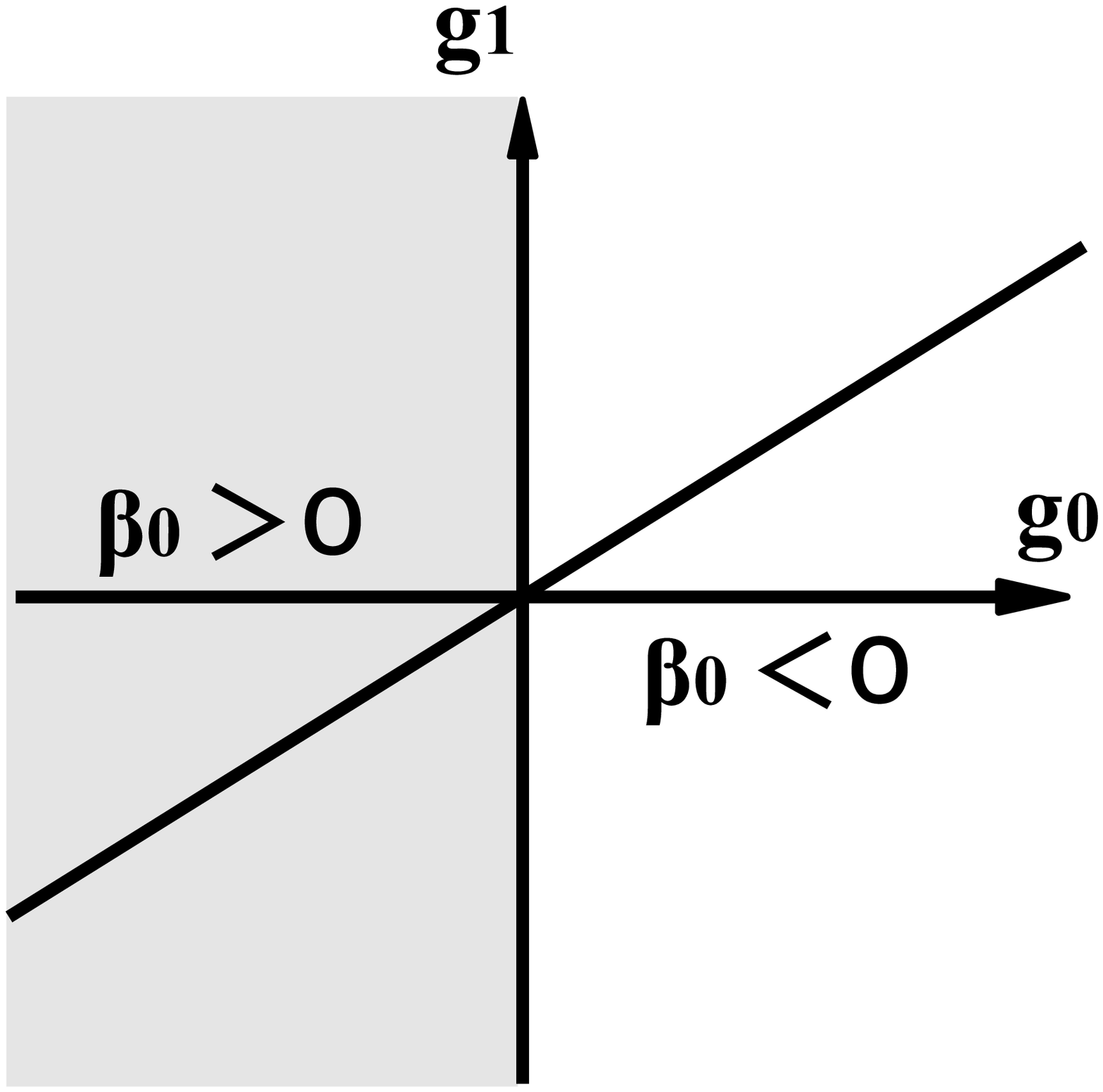} \\ \vspace{-2cm} Fig.3c:
 $\beta_0>0$ in the grayed region. The oblique line represents $g_1=\frac{\lambda^{-2}}{32m^2}g_1$.
\end{center}
\end{minipage}
\end{center}

\noindent
(ii) $\lambda^{-1} = 0$ (Fig.4a-4c) \\
$\beta_0 = 0$. For $\lambda^{-1}$, \\
$\bullet$ \ $\beta_{\lambda^{-1}} > 0$, if \ $\{ g_0 > 0 \ , \ g_1 > 0 \}$
\ or \ $\{ g_0 < 0 \ , \ g_1 < 0 \}$, \\
$\bullet$ \ $\beta_{\lambda^{-1}} < 0$, if \ $\{ g_0 > 0 \ , \ g_1 < 0 \}$ \
or \ $\{ g_0 < 0 \ , \ g_1 > 0 \}$, \\
$\bullet$ \ $\beta_{\lambda^{-1}} = 0$, if \ $\{ g_0 = 0 \}$ \ or \ $\{
g_1 = 0 \}$.

\begin{center}
\begin{minipage}{5cm}
\vspace{-1cm}
\begin{center}
\includegraphics[width=5cm,clip]{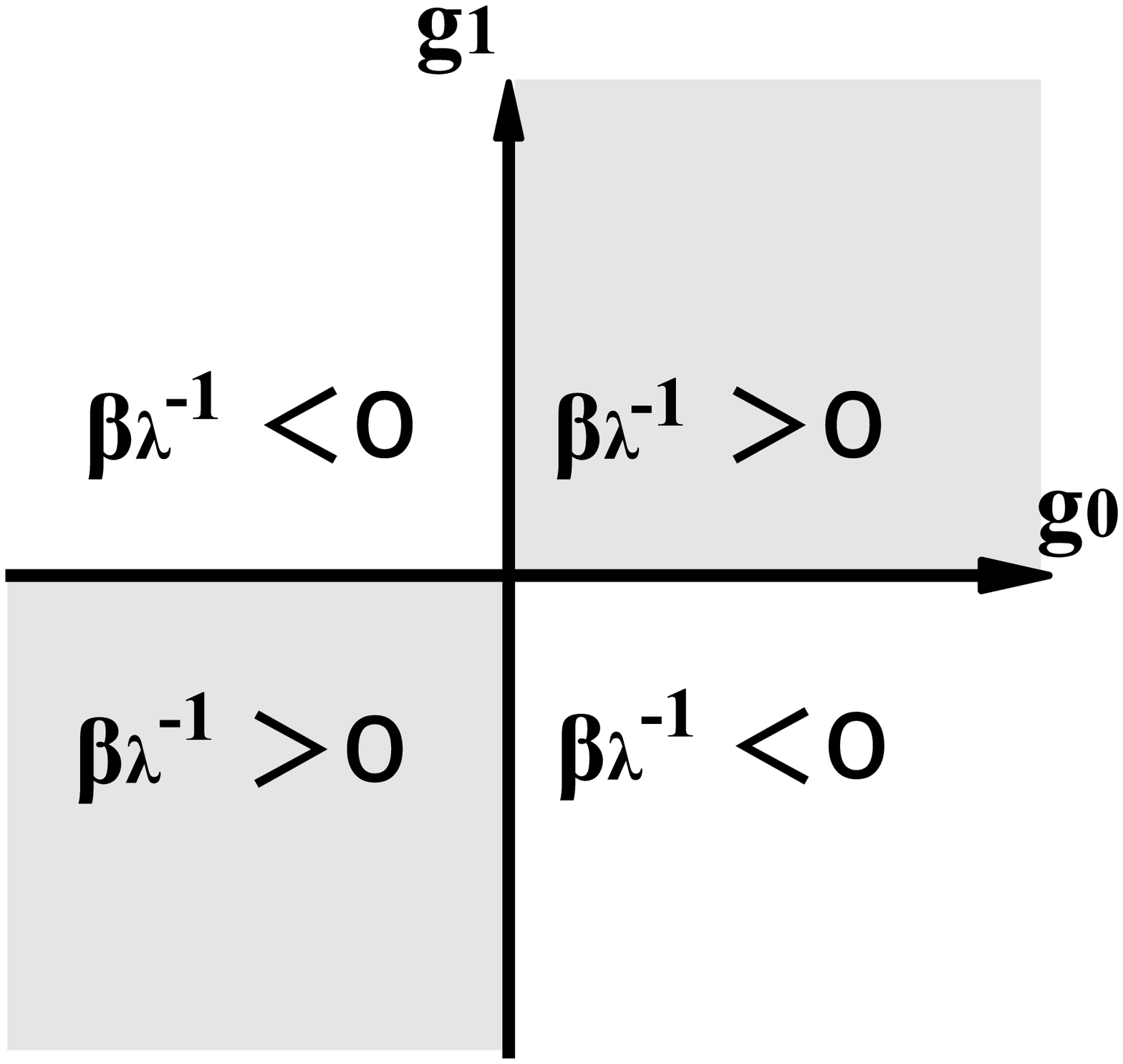} \\ \vspace{-2cm} Fig.4a:
 $\beta_{\lambda^{-1}}>0$ in the grayed region.
\end{center}
\end{minipage} \ \
\begin{minipage}{5cm}
\vspace{-1cm}
\begin{center}
\includegraphics[width=5cm,clip]{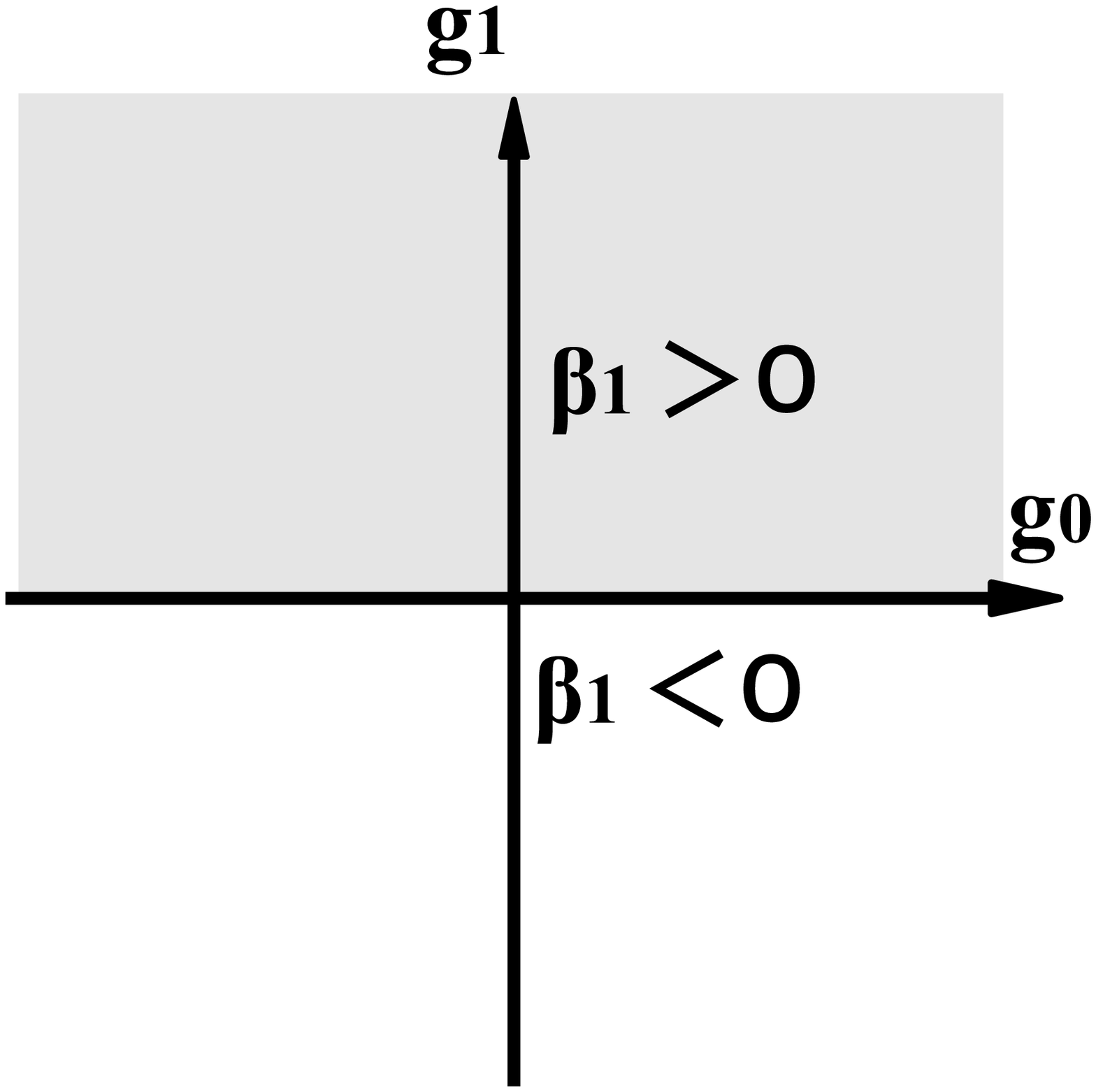} \\ \vspace{-2cm} Fig.4b:
 $\beta_{\tilde 1}>0$ in the grayed region.
\end{center}
\end{minipage} \ \
\begin{minipage}{5cm}
\vspace{-1cm}
\begin{center}
\includegraphics[width=5cm,clip]{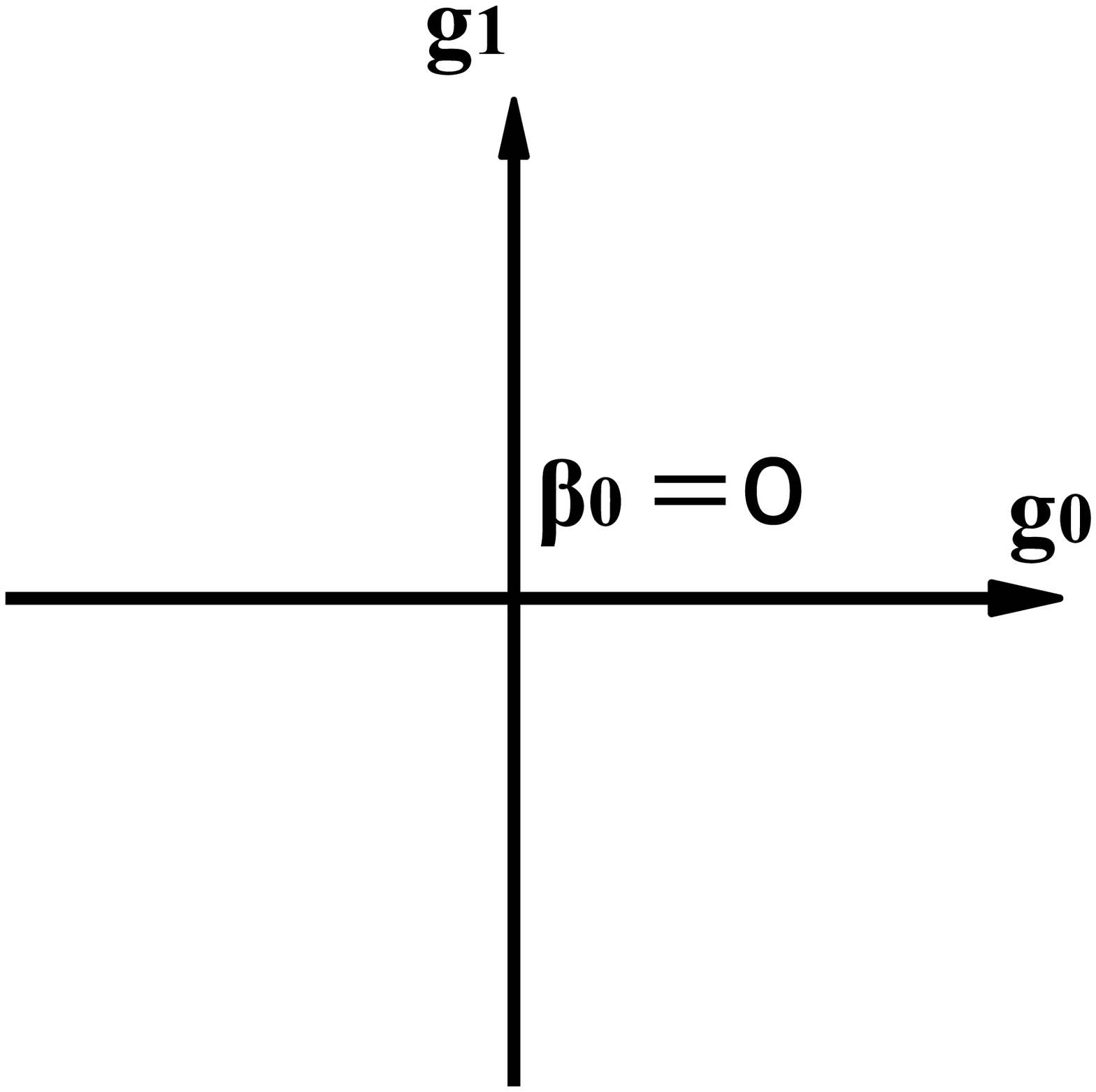} \\ \vspace{-2cm} Fig.4c: $\beta_0=0$ in
 the whole (perturbative) region.
\end{center}
\end{minipage}
\end{center}

\noindent
(iii) $\lambda^{-1} < 0$ (Fig.5a-5c) \\
For $g_0$, $g_0 \rightarrow 0$ in the IR limit. For $\lambda^{-1}$, \\
$\bullet$ \ $\lambda^{-1} \rightarrow 0$ in UV, if \ $\{ g_0 > 0 \ , \
g_1 > \frac{\lambda^{-2}}{32 m^2} g_0 \}$ \ or \ $\{ g_0 <
0 \ , \ g_1 < \frac{\lambda^{-2}}{32 m^2} g_0 \}$, \\
$\bullet$ \ $\lambda^{-1} \rightarrow 0$ in IR, if \ $\{ g_0 > 0 \ , \
g_1 < \frac{\lambda^{-2}}{32 m^2} g_0 \}$ \ or \ $\{ g_0 <
0 \ , \ g_1 > \frac{\lambda^{-2}}{32 m^2} g_0 \}$.

\begin{center}
\begin{minipage}{5cm}
\vspace{-1cm}
\begin{center}
\includegraphics[width=5cm,clip]{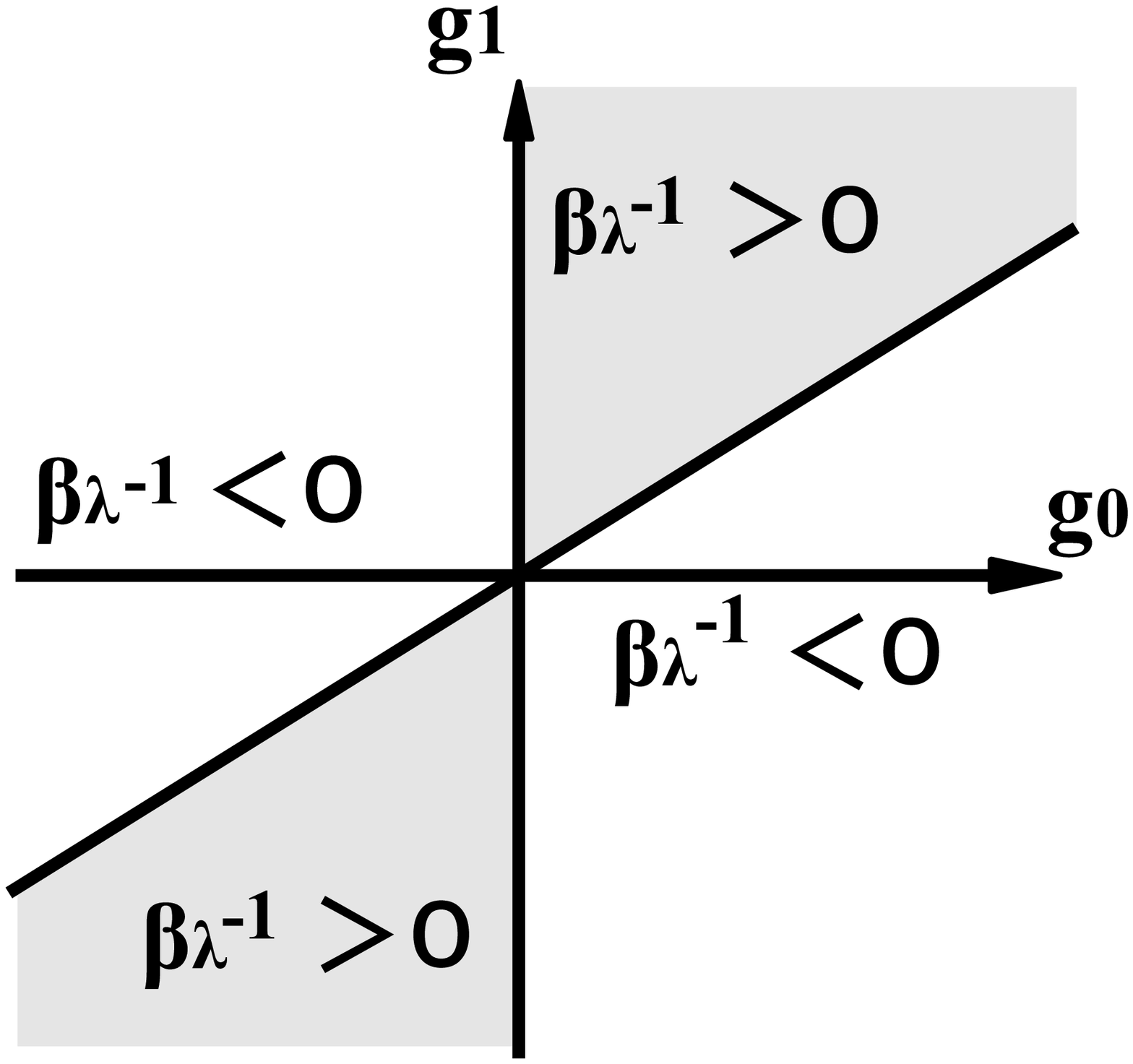} \\ \vspace{-2cm} Fig.5a:
 $\beta_{\lambda^{-1}}>0$ in the grayed region. The oblique line
 represents $g_1=\frac{\lambda^{-2}}{32m^2}g_0$.
\end{center}
\end{minipage} \ \ 
\begin{minipage}{5cm}
\vspace{-1cm}
\begin{center}
\includegraphics[width=5cm,clip]{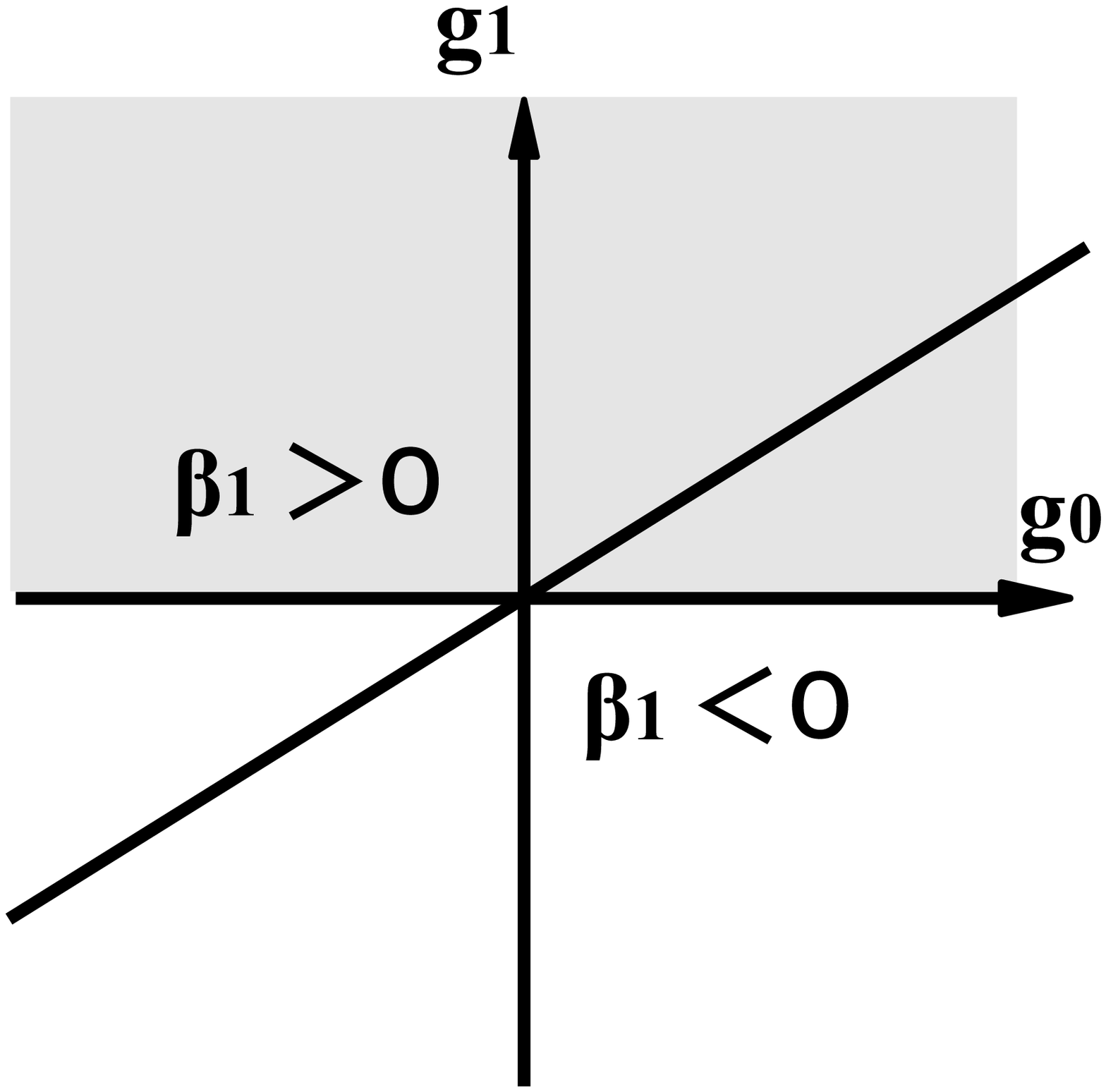} \\ \vspace{-2cm} Fig.5b:
 $\beta_{\tilde 1}>0$ in the grayed region. The oblique line represents $g_1=\frac{\lambda^{-2}}{32m^2}g_0$.
\end{center}
\end{minipage} \ \ 
\begin{minipage}{5cm}
\vspace{-1cm}
\begin{center}
\includegraphics[width=5cm,clip]{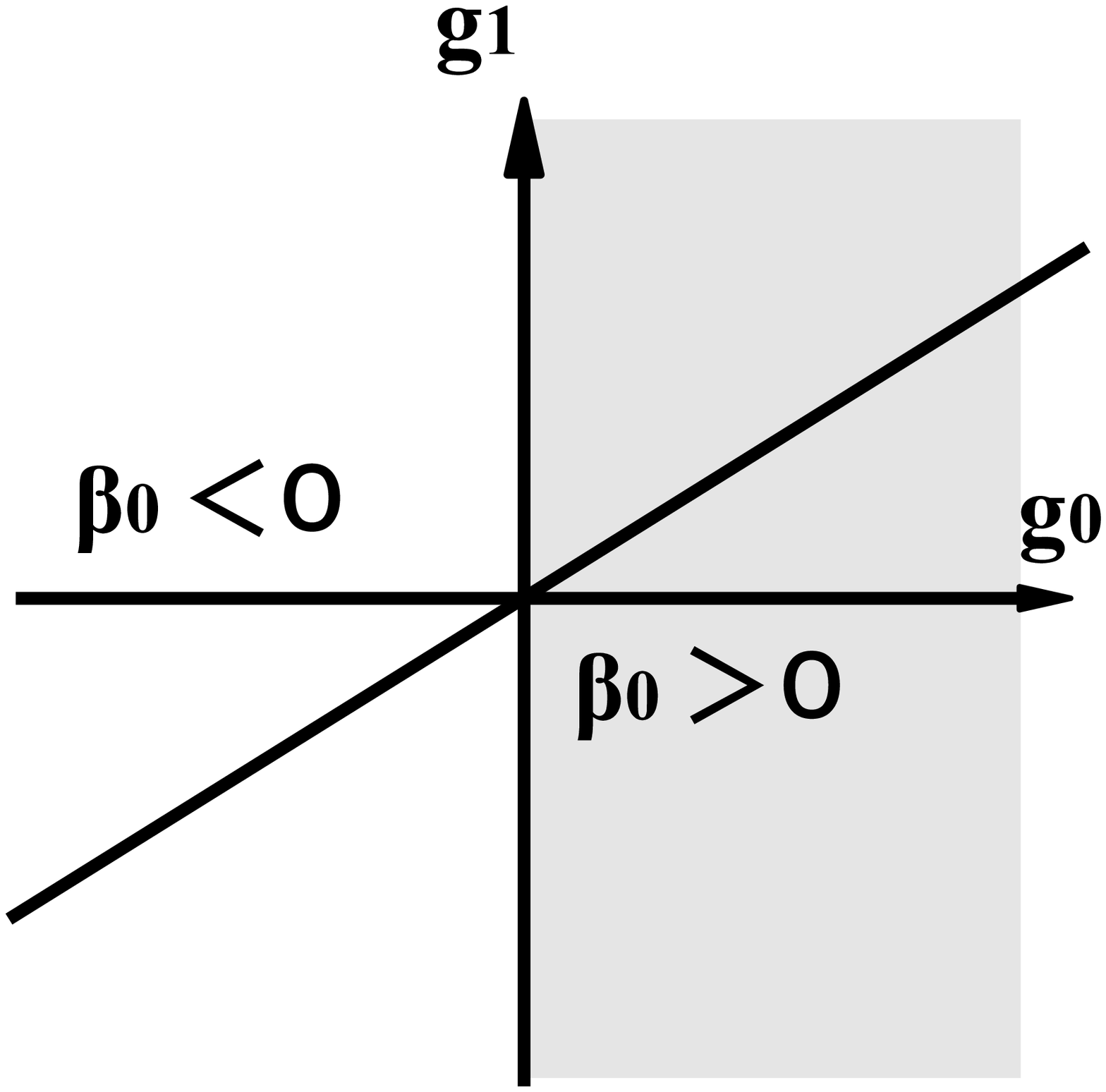} \\ \vspace{-2cm} Fig.5c: $\beta_0>0$ in the grayed region. The oblique line represents $g_1=\frac{\lambda^{-2}}{32m^2}g_0$.  
\end{center}
\end{minipage}
\end{center}

{}From the above results, full SUSY is realized in the IR limit if
the RG flow starts from
\begin{eqnarray}
 &\bullet& \{ \lambda^{-1} >0 \ , \ g_0 > 0 \ , \ g_1 > \frac{\lambda^{-2}}{32 m^2} g_0 \} \ , \label{IRSUSY1} \\
\mbox{or} & & \nonumber \\
 &\bullet& \{ \lambda^{-1} >0 \ , \ g_0 < 0 \ , \ g_1 < \frac{\lambda^{-2}}{32 m^2} g_0 \} \ , \label{IRSUSY2} \\
\mbox{or} & & \nonumber \\
&\bullet& \{ \lambda^{-1} <0 \ , \ g_0 > 0 \ , \ g_1 < \frac{\lambda^{-2}}{32 m^2} g_0 \} \ , \label{IRSUSY3} \\
\mbox{or} & & \nonumber \\
&\bullet& \{ \lambda^{-1} <0 \ , \ g_0 > 0 \ , \ g_1 < \frac{\lambda^{-2}}{32 m^2} g_0 \} \ .
\label{IRSUSY4} 
\end{eqnarray}

By using the relations (\ref{g-g*}) and $\lambda_* = \lambda$, the above
argument can be rewritten in terms of the non(anti-)commutative field 
theory and the super-$*$ symmetry.

\noindent
(i)$\lambda_*^{-1} > 0$ (Fig.6a-6c) \\
For $g_{0*}$, $g_{0*} \rightarrow 0$ in the UV limit. For $\lambda_*^{-1}$, \\
$\bullet$ \ $\lambda_*^{-1} \rightarrow 0$ in IR, \\ 
\ \ \ \ if \ $\{ g_{0*} > 0 \ , \ g_{1*} > ( \frac{1}{4} det P +
\frac{\lambda_*^{-2}}{32 m^2} ) g_{0*} \}$ \ or \ $\{ g_{0*} < 0 \ , \
g_{1*} < ( \frac{1}{4} det P + \frac{\lambda_*^{-2}}{32 m^2} ) g_{0*} \}$, \\
$\bullet$ \ $\lambda_*^{-1} \rightarrow 0$ in UV, \\
\ \ \ \ if \ $\{ g_{0*} > 0 \
, \ g_{1*} < ( \frac{1}{4} det P +
\frac{\lambda^{-2}}{32 m^2} ) g_{0*} \}$ \ or \ $\{ g_{0*} <
0 \ , \ g_{1*} > ( \frac{1}{4} det P + \frac{\lambda_*^{-2}}{32 m^2} )
g_{0*} \}$.

\begin{center}
\begin{minipage}{5cm}
\vspace{-1cm}
\begin{center}
\includegraphics[width=5cm,clip]{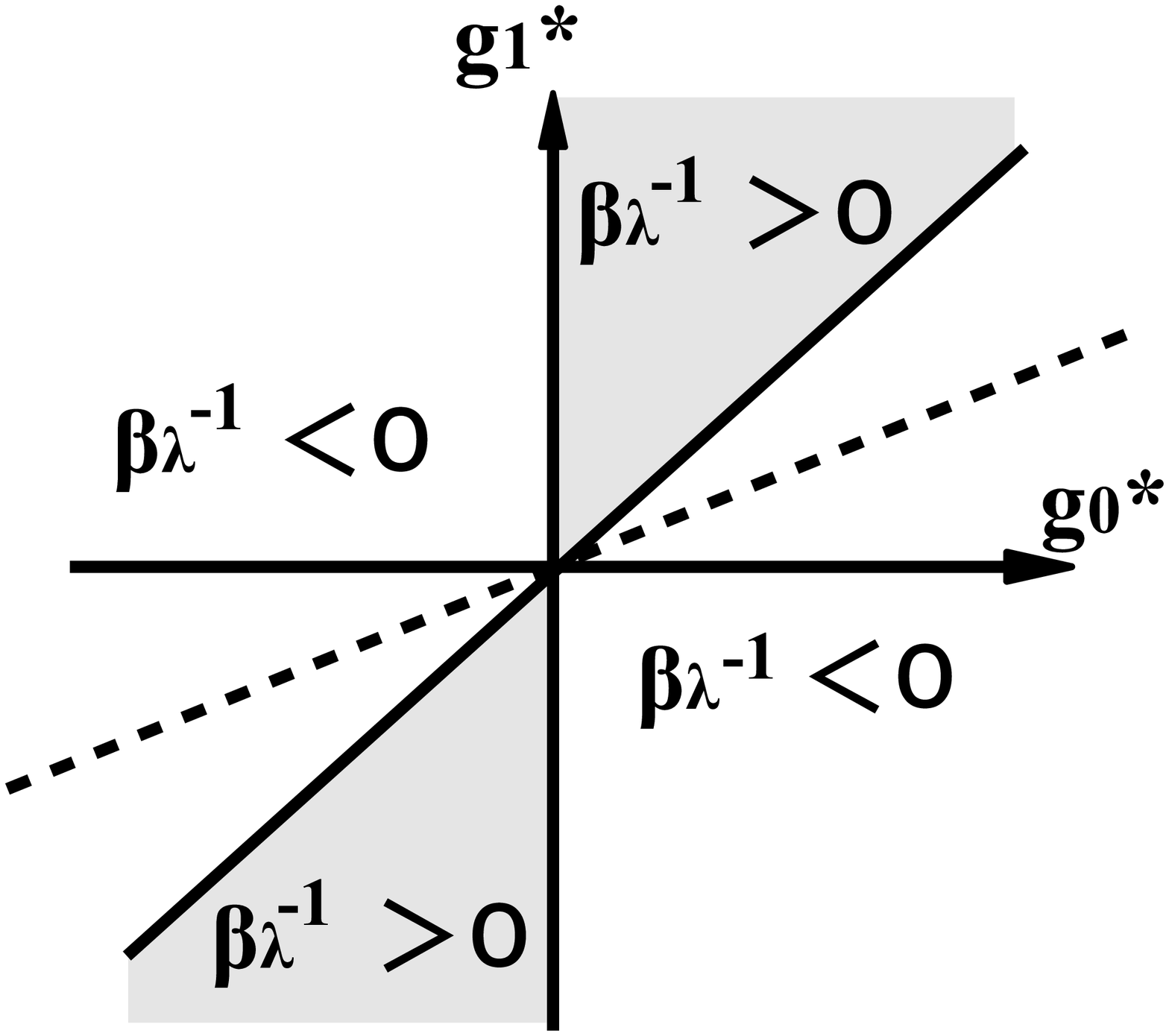} \\ \vspace{-2cm} Fig.6a:
 $\beta_{\lambda^{-1}}>0$ in the grayed region. The dash oblique line
 represents $g_{1*}=\frac{1}{4} det P g_{0*}$ and the solid oblique one
 does $g_{1*} = ( \frac{1}{4} det P + \frac{\lambda_*^{-2}}{32 m^2} )
g_{0*}$.
\end{center}
\end{minipage} \ \
\begin{minipage}{5cm}
\vspace{-1cm}
\begin{center}
\includegraphics[width=5cm,clip]{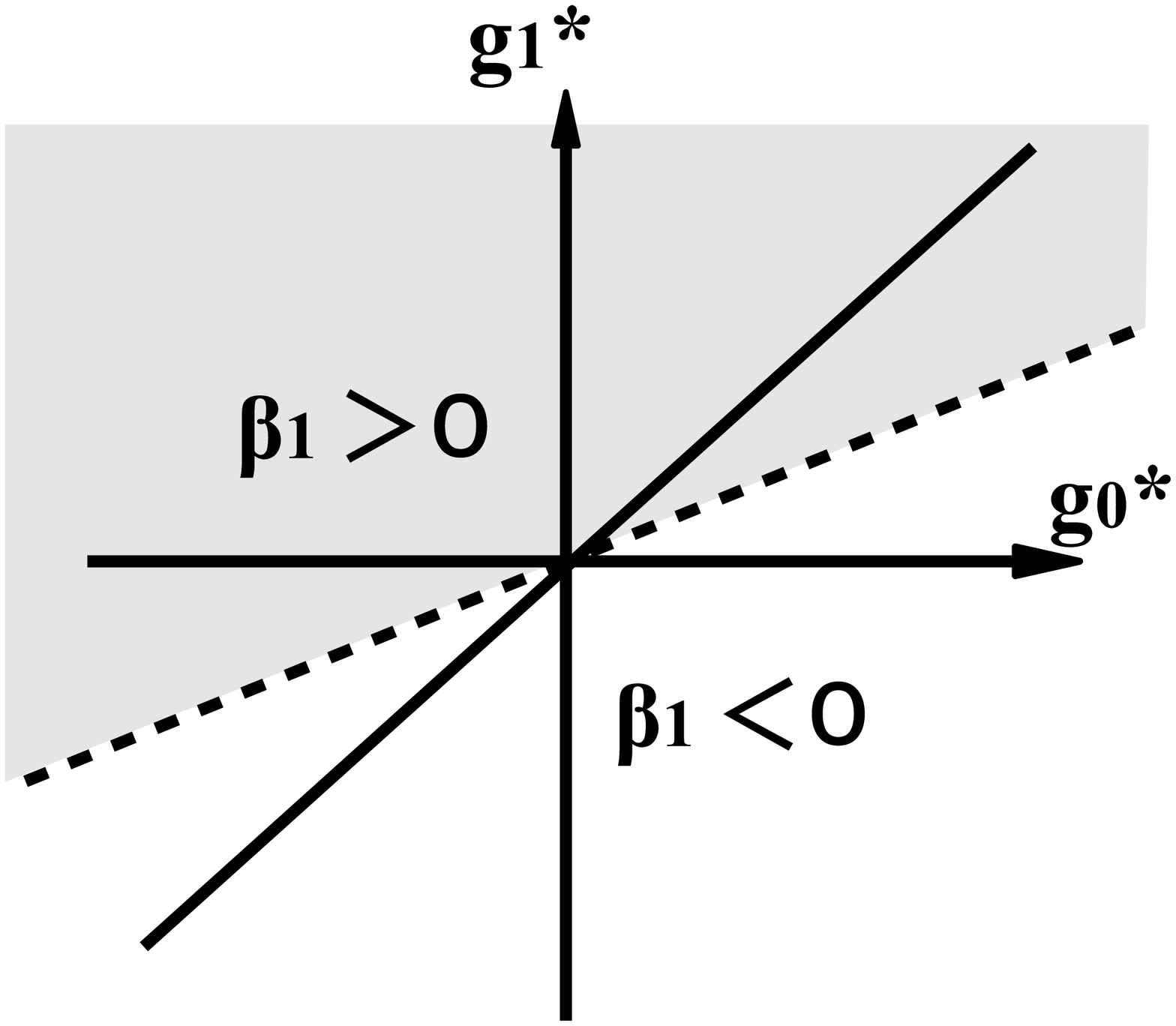} \\ \vspace{-2cm} Fig.6b:
 $\beta_{\tilde 1}>0$ in the grayed region. The dash oblique line represents
 $g_{1*}=\frac{1}{4} det P g_{0*}$ and the solid oblique one does
 $g_{1*} = ( \frac{1}{4} det P + \frac{\lambda_*^{-2}}{32 m^2} ) g_{0*}$.
\end{center}
\end{minipage} \ \
\begin{minipage}{5cm}
\vspace{-1cm}
\begin{center}
\includegraphics[width=5cm,clip]{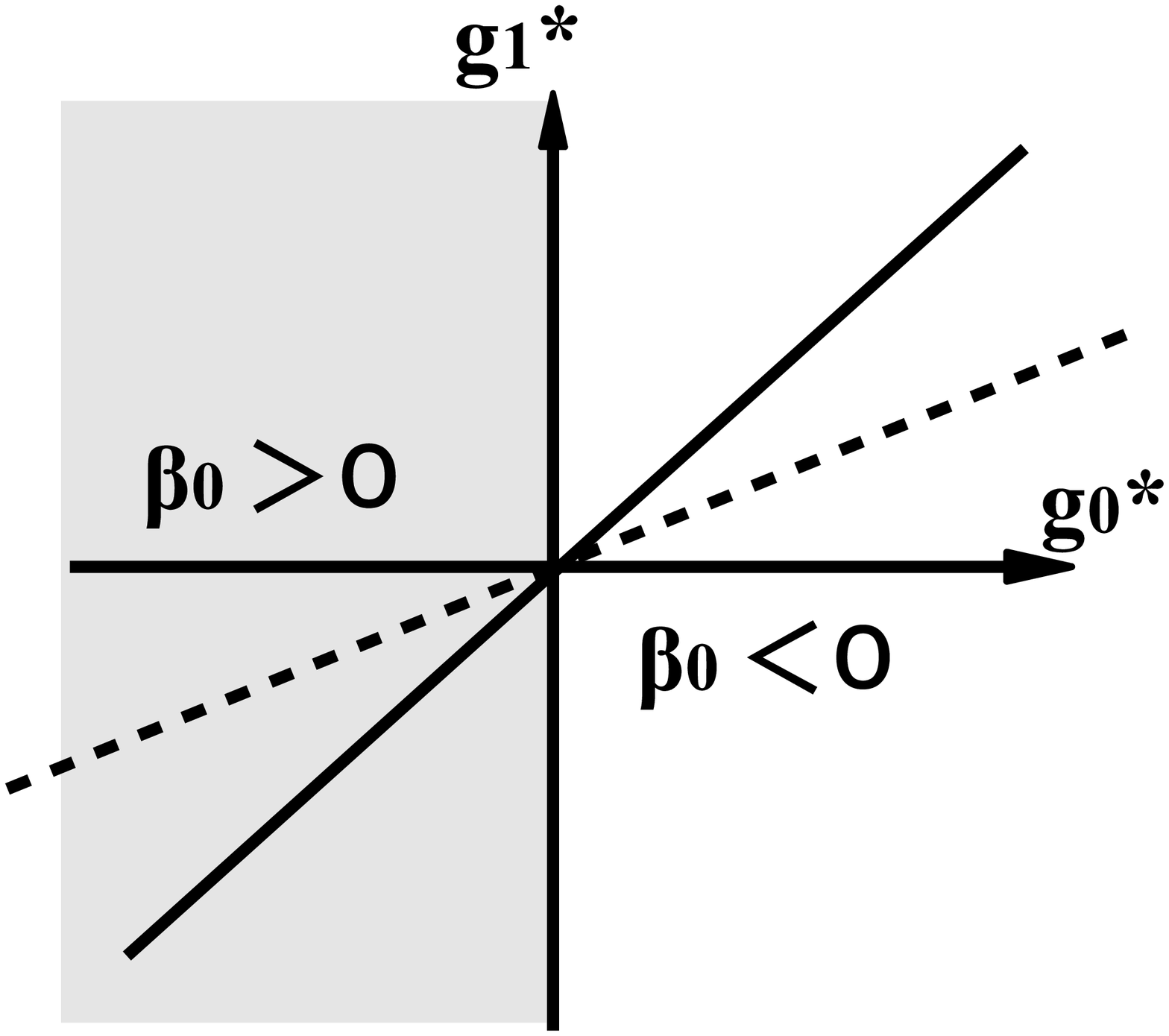} \\ \vspace{-2cm} Fig.6c:
 $\beta_0>0$ in the grayed region. The dash oblique line represents
 $g_{1*}=\frac{1}{4} det P g_{0*}$ and the solid oblique one does
 $g_{1*} = ( \frac{1}{4} det P + \frac{\lambda_*^{-2}}{32 m^2} ) g_{0*}$.
\end{center}
\end{minipage}
\end{center}

\noindent
(ii) $\lambda^{-1} = 0$ (Fig.7a-7c) \\
$\beta_0 = 0$. For $\lambda^{-1}$, \\
$\bullet$ \ $\beta_{\lambda^{-1}} > 0$, if \ $\{ g_{0*} > 0 \ , \ g_{1*}
> \frac{1}{4} det P g_{0*} \}$ \ or \ $\{ g_{0*} <
0 \ , \ g_{1*} < \frac{1}{4} det P g_{0*} \}$, \\
$\bullet$ \ $\beta_{\lambda^{-1}} < 0$, if \ $\{ g_{0*} > 0 \ , \
g_{1*} < \frac{1}{4} det P g_{0*} \}$ \ or \ $\{ g_{0*}
< 0 \ , \ g_{1*} > \frac{1}{4} det P g_{0*} \}$, \\
$\bullet$ \ $\beta_{\lambda^{-1}} = 0$, if \ $g_{0*} = 0$ \ or \ $g_{1*} = \frac{1}{4} det P g_{0*}$.

\begin{center}
\begin{minipage}{5cm}
\vspace{-1cm}
\begin{center}
\includegraphics[width=5cm,clip]{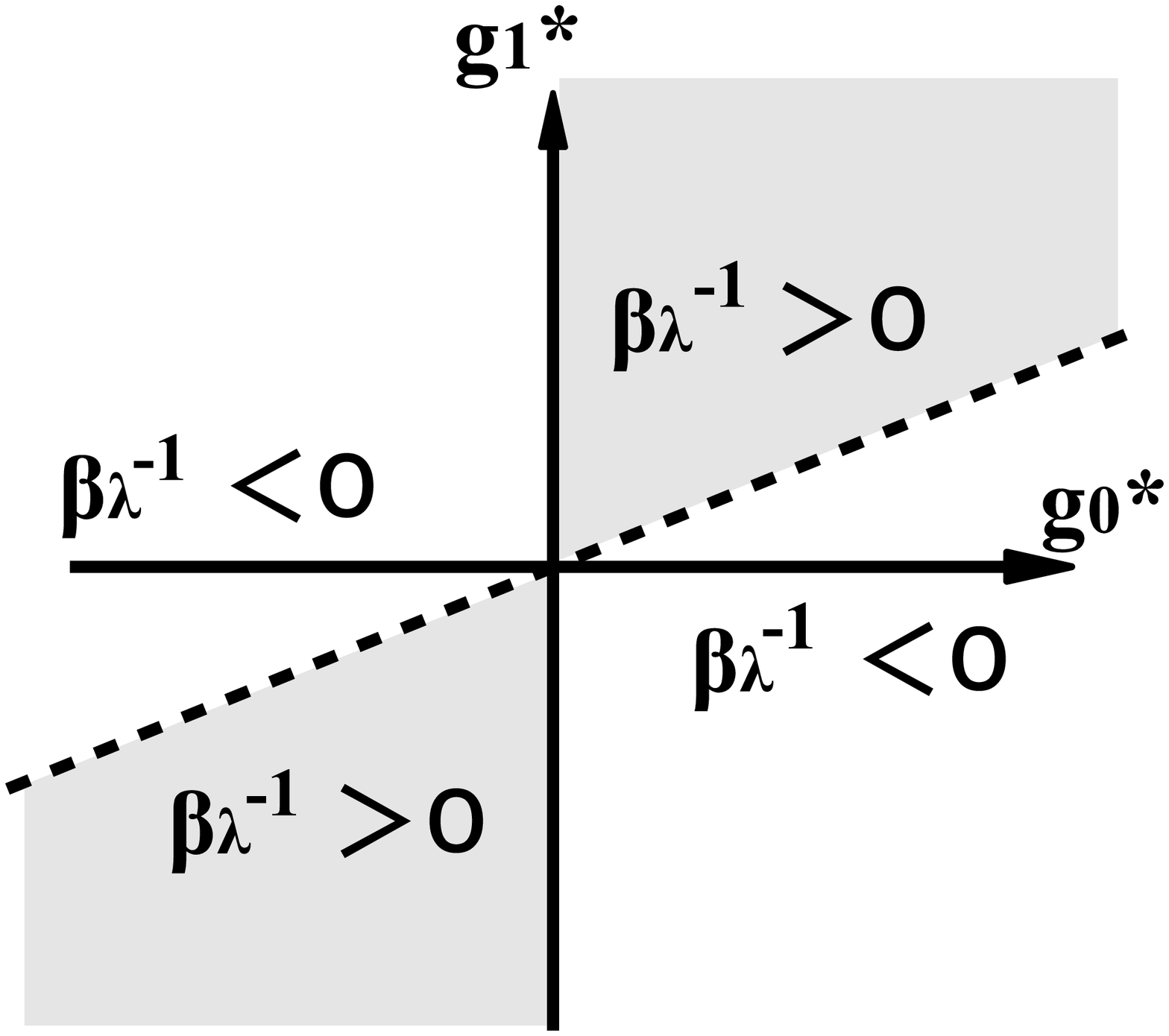} \\ \vspace{-2cm} Fig.7a: $\beta_{\lambda^{-1}}>0$ in the grayed region. The dash oblique line represents
 $g_{1*}=\frac{1}{4} det P g_{0*}$.
\end{center}
\end{minipage} \ \
\begin{minipage}{5cm}
\vspace{-1cm}
\begin{center}
\includegraphics[width=5cm,clip]{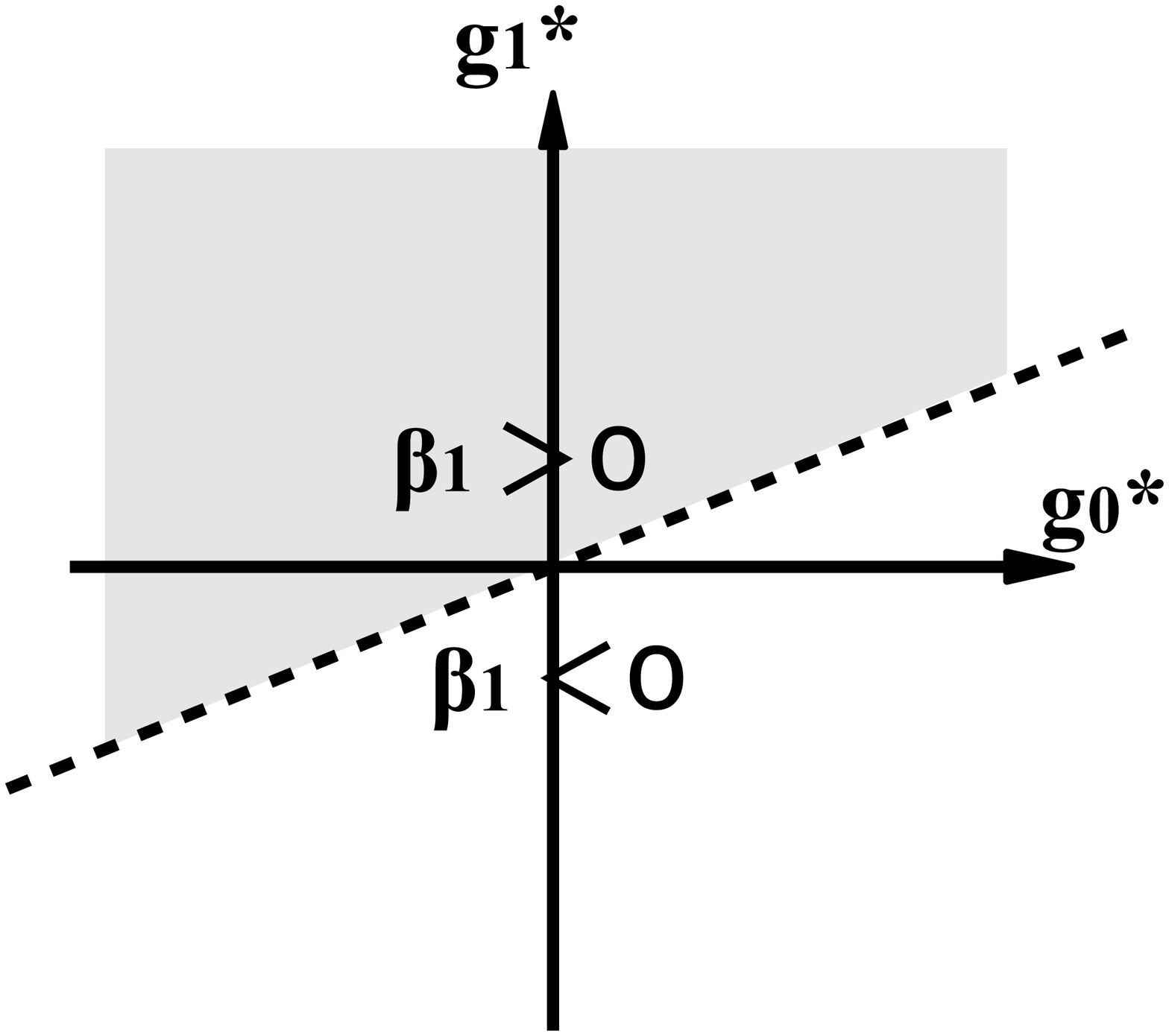} \\ \vspace{-2cm} Fig.7b:
 $\beta_{\tilde 1}>0$ in the grayed region. The dash oblique line represents
 $g_{1*}=\frac{1}{4} det P g_{0*}$.
\end{center}
\end{minipage} \ \
\begin{minipage}{5cm}
\vspace{-1cm}
\begin{center}
\includegraphics[width=5cm,clip]{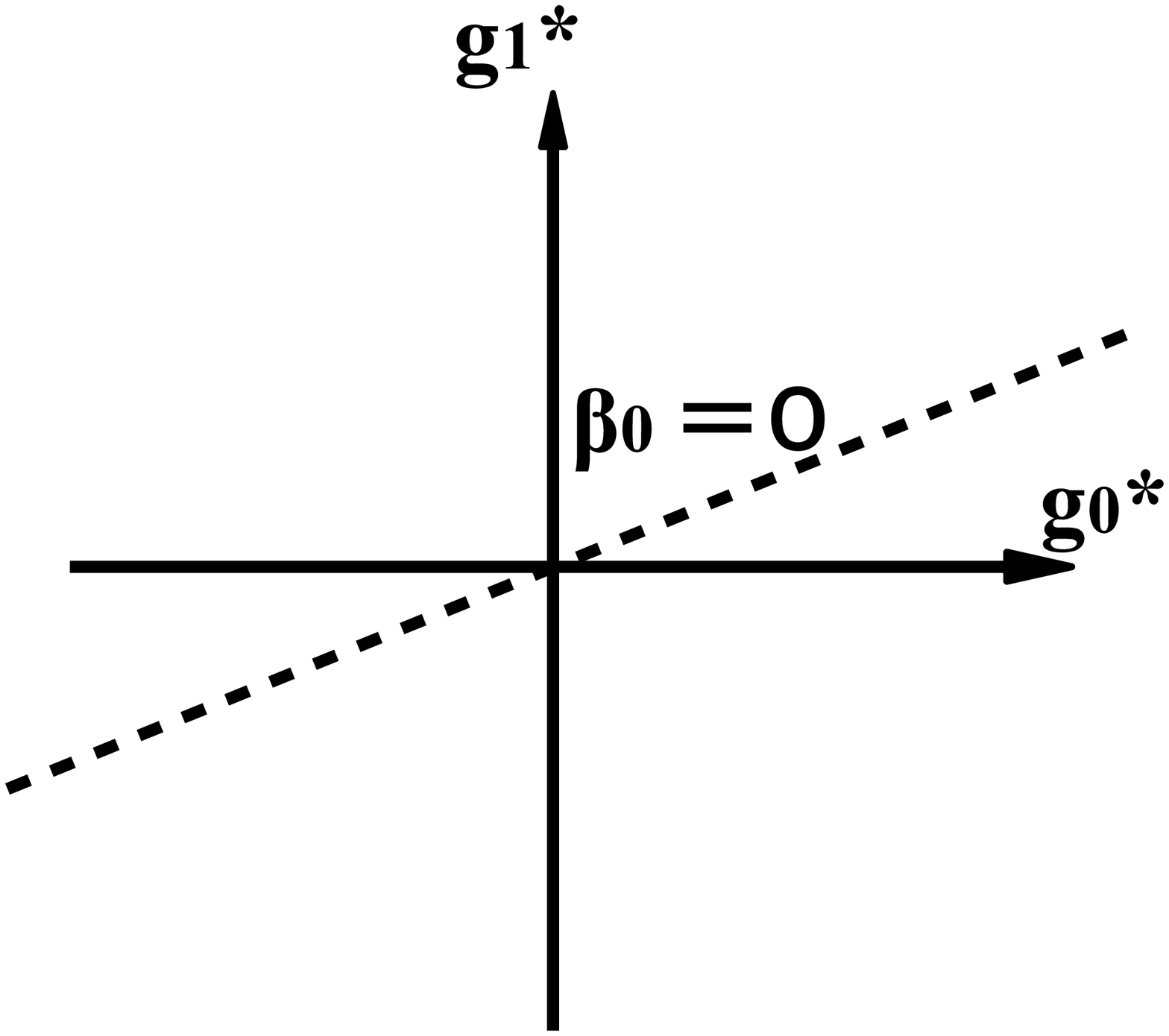} \\ \vspace{-2cm} Fig.7c:
 $\beta_0=0$ in the whole (perturbative) region. The dash oblique line
 represents $g_{1*}=\frac{1}{4} det P g_{0*}$.
\end{center}
\end{minipage}
\end{center}

\noindent
(iii) $\lambda_*^{-1} < 0$ (Fig.8a-8c) \\
For $g_{0*}$, $g_{0*} \rightarrow 0$ in the IR limit. For $\lambda_*^{-1}$, \\
$\bullet$ \ $\lambda_*^{-1} \rightarrow 0$ in UV, \\
\ \ \ \ if \ $\{ g_{0*} > 0 \ ,
\ g_{1*} > ( \frac{1}{4} det P +
\frac{\lambda_*^{-2}}{32 m^2} ) g_{0*} \}$ \ or \ $\{ g_{0*} <
0 \ , \ g_{1*} < ( \frac{1}{4} det P + \frac{\lambda_*^{-2}}{32 m^2} )
g_{0*} \}$, \\
$\bullet$ \ $\lambda_*^{-1} \rightarrow 0$ in IR, \\
\ \ \ \ if \ $\{ g_{0*} > 0 \ ,
\ g_{1*} < ( \frac{1}{4} det P +
\frac{\lambda_*^{-2}}{32 m^2} ) g_{0*} \}$ \ or \ $\{ g_{0*} <
0 \ , \ g_{1*} > ( \frac{1}{4} det P + \frac{\lambda_*^{-2}}{32 m^2} )
g_{0*} \}$. 

\begin{center}
\begin{minipage}{5cm}
\vspace{-1cm}
\begin{center}
\includegraphics[width=5cm,clip]{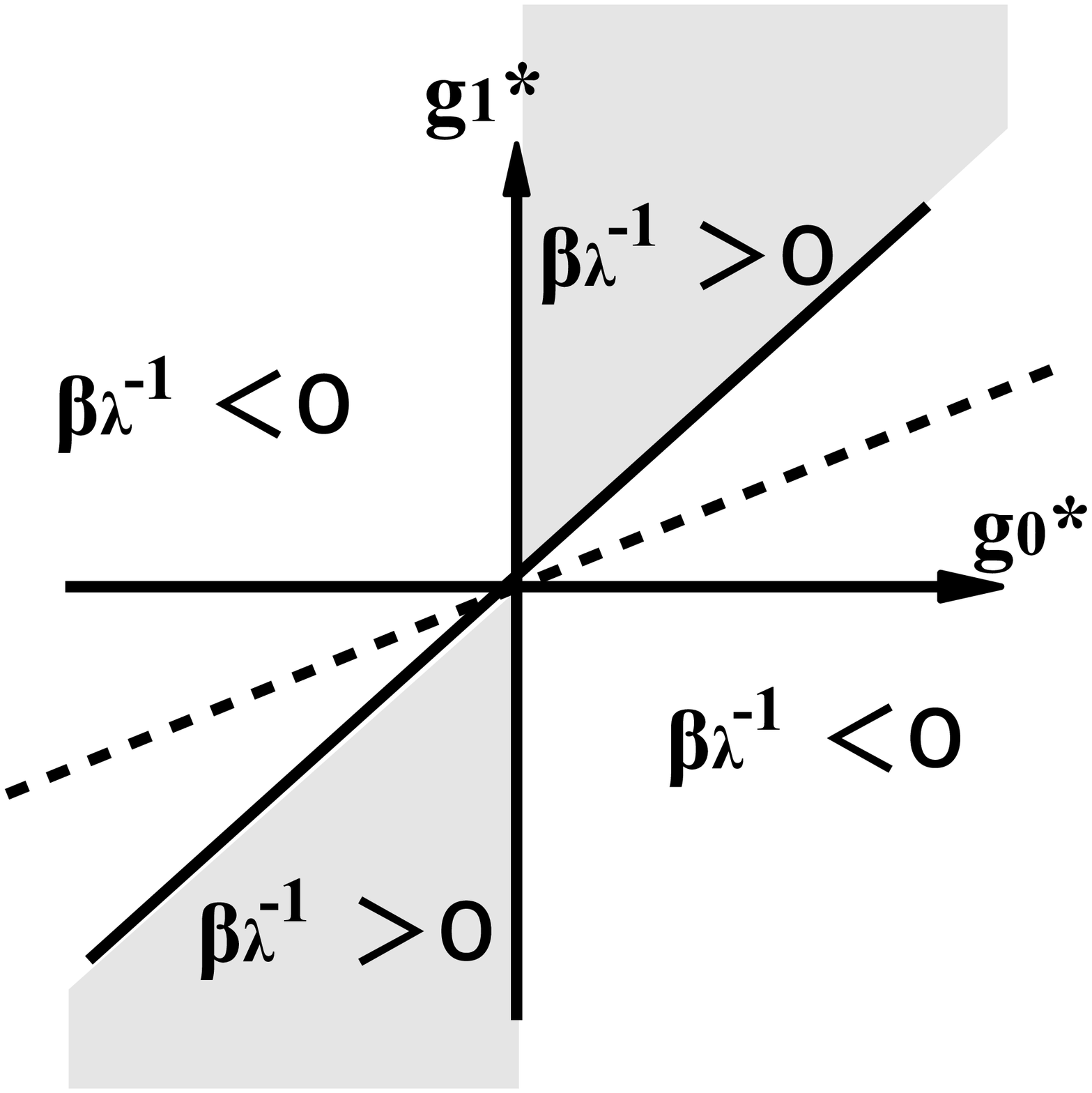} \\ \vspace{-2cm} Fig.8a:
 $\beta_{\lambda^{-1}}>0$ in the grayed region. The dash oblique line
 represents $g_{1*}=\frac{1}{4} det P g_{0*}$ and the solid oblique one
 does $g_{1*} = ( \frac{1}{4} det P + \frac{\lambda_*^{-2}}{32 m^2} ) g_{0*}$.
\end{center}
\end{minipage} \ \
\begin{minipage}{5cm}
\vspace{-1cm}
\begin{center}
\includegraphics[width=5cm,clip]{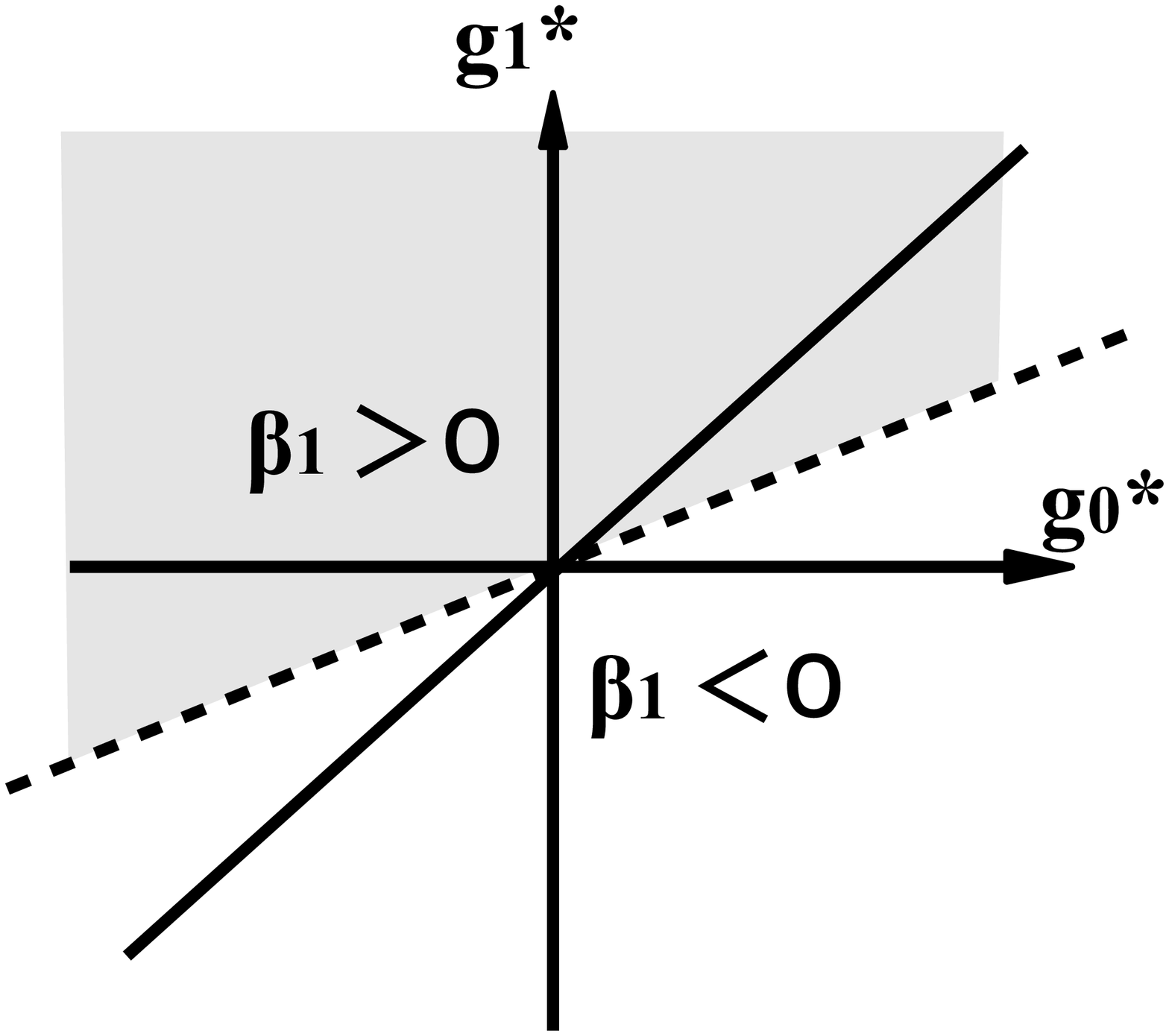} \\ \vspace{-2cm} Fig.8b:
  $\beta_{\tilde 1}>0$ in the grayed region. The dash oblique line
 represents $g_{1*}=\frac{1}{4} det P g_{0*}$ and the solid oblique one
 does $g_{1*} = ( \frac{1}{4} det P + \frac{\lambda_*^{-2}}{32 m^2} ) g_{0*}$.
\end{center}
\end{minipage} \ \
\begin{minipage}{5cm}
\vspace{-1cm}
\begin{center}
\includegraphics[width=5cm,clip]{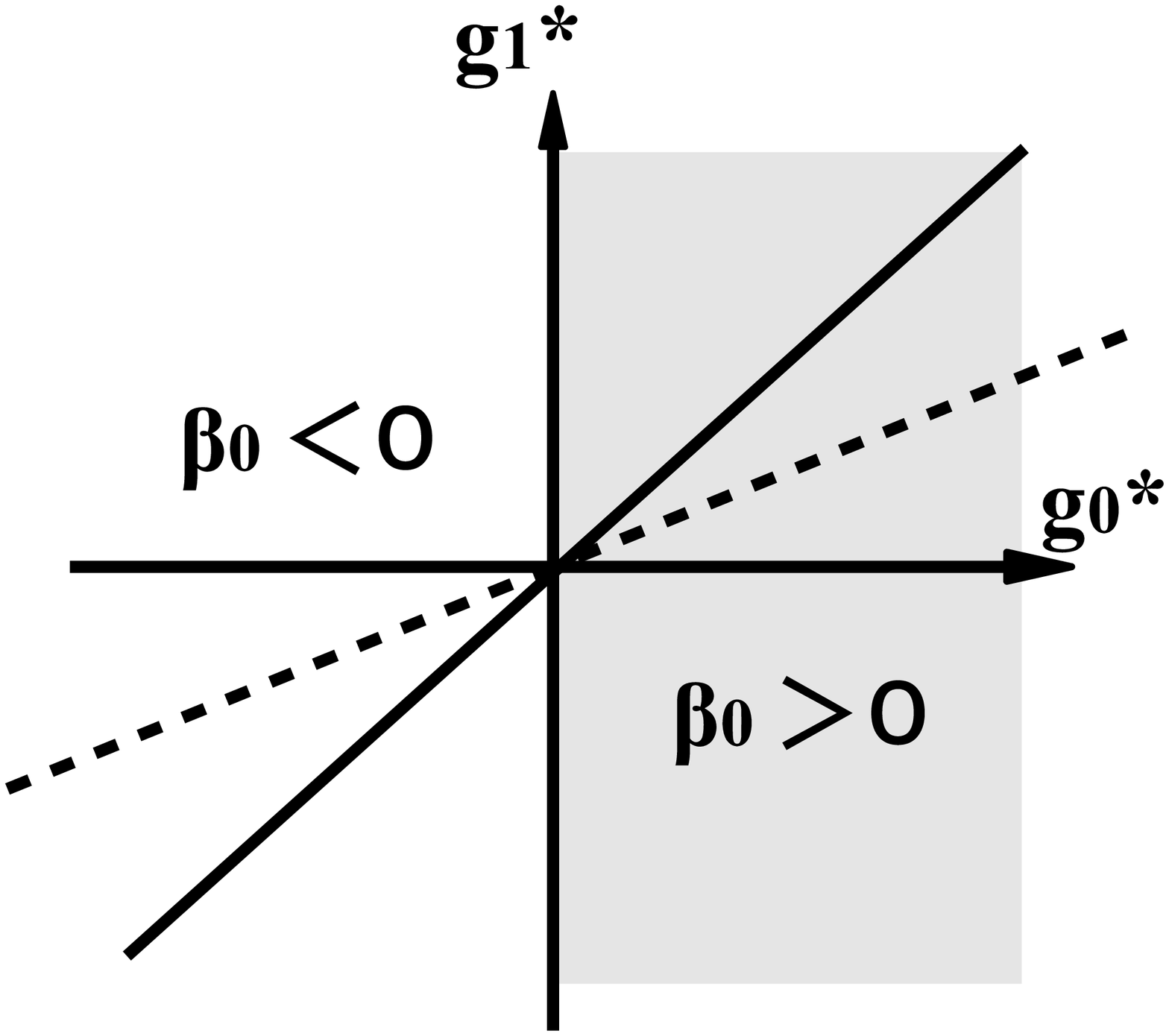} \\ \vspace{-2cm} Fig.8c:
 $\beta_0>0$ in the grayed region. The dash oblique line represents
 $g_{1*}=\frac{1}{4} det P g_{0*}$ and the solid oblique one does
 $g_{1*} = ( \frac{1}{4} det P + \frac{\lambda_*^{-2}}{32 m^2} ) g_{0*}$.
\end{center}
\end{minipage}
\end{center}

{}From the above results, the super-$*$ symmetry is realized in the IR limit if
the RG flow starts from
\begin{eqnarray}
 &\bullet& \{ \lambda_*^{-1} >0 \ , \ g_{0*} > 0 \ , \ g_{1*} > ( \frac{1}{4} det P + \frac{\lambda_*^{-2}}{32 m^2} ) g_{0*} \} \ , \label{IRSUSY1*} \\
\mbox{or} & & \nonumber \\
 &\bullet& \{ \lambda_*^{-1} >0 \ , \ g_{0*} < 0 \ , \ g_{1*} < ( \frac{1}{4} det P + \frac{\lambda_*^{-2}}{32 m^2} ) g_{0*} \} \ , \label{IRSUSY2*} \\
\mbox{or} & & \nonumber \\
&\bullet& \{ \lambda_*^{-1} <0 \ , \ g_{0*} > 0 \ , \ g_{1*} < ( \frac{1}{4} det P + \frac{\lambda_*^{-2}}{32 m^2} ) g_{0*} \} \ , \label{IRSUSY3*} \\
\mbox{or} & & \nonumber \\
&\bullet& \{ \lambda_*^{-1} <0 \ , \ g_{0*} > 0 \ , \ g_{1*} < ( \frac{1}{4} det P + \frac{\lambda_*^{-2}}{32 m^2} ) g_{0*} \} \ .
\label{IRSUSY4*} 
\end{eqnarray}

\section{Conclusions}
In this article we gave the general prescription to construct full SUSY
lagrangians in non(anti-)commutative superspaces, which is relevant
for both the SUSY $*$-deformed superspace and the non-SUSY
$\star$-deformed superspace. 

We investigated quantum effects at the 1-loop level for the deformed
$\Phi_*^3$ Wess-Zumino model with the $\Phi D^2 \Phi$ proportional term.
We found that this term yields a divergent tadpole graph and
the renormalization of this causes the wave
function renormalization.
The wave function renormalization gave the non-trivial $\beta$-
functions.

From the obtained $\beta$-functions, we found
the conditions on the parameters $(g_0,g_1,\lambda^{-1})$ to recover full SUSY in the IR limit. We also rewrote the conditions into ones on
$(g_{o*},g_{1*},\lambda_*^{-1})$, the parameters of the
non(anti-)commutative field theories.

\end{document}